\documentclass[12pt]{article}

\usepackage{comment}
\usepackage{pdflscape}
\usepackage{amsmath}
\usepackage{graphicx,psfrag,epsf}
\usepackage{enumerate}
\usepackage{biblatex} % 'sorting=nty' is the default

\usepackage{url} % not crucial - just used below for the URL 
\usepackage[utf8]{inputenc} % allow utf-8 input
\usepackage[T1]{fontenc}    % use 8-bit T1 fonts
\usepackage{hyperref}       % hyperlinks

\usepackage{booktabs}       % professional-quality tables
\usepackage{amsfonts}       % blackboard math symbols
\usepackage{nicefrac}       % compact symbols for 1/2, etc.
\usepackage{microtype}      % microtypography
\usepackage{lipsum}
\usepackage{fancyhdr}       % header
\usepackage{graphicx}       % graphics
\graphicspath{{media/}}     % organize your images and other figures under media/ folder
\usepackage{amsthm}
\usepackage{newtxmath}
\usepackage{booktabs}
\usepackage{float}  
\usepackage{subfigure}
\usepackage{subcaption}
\usepackage{xcolor}
\usepackage{graphicx}
\newtheorem{theorem}{Theorem}

\usepackage{xr}

\makeatletter
\newcommand*{\addFileDependency}[1]{% argument=file name and extension
  \typeout{(#1)}
  \@addtofilelist{#1}
  \IfFileExists{#1}{}{\typeout{No file #1.}}
}
\makeatother

\newcommand*{\myexternaldocument}[1]{%
    \externaldocument{#1}%
    \addFileDependency{#1.tex}%
    \addFileDependency{#1.aux}%
}

\myexternaldocument{SM}

\newcommand{\blind}{1}

\begin{document}

\def\spacingset#1{\renewcommand{\baselinestretch}%
{#1}\small\normalsize} \spacingset{1}

%%%%%%%%%%%%%%%%%%%%%%%%%%%%%%%%%%%%%%%%%%%%%%%%%%%%%%%%%%%%%%%%%%%%%%%%%%%%%%

\if1\blind
{
  \title{\bf On the interplay between prior weight and variance of the robustification component in Robust Mixture Prior Bayesian Dynamic Borrowing approach}
  \author{Marco Ratta \hspace{.2cm}\\
    Department of Mathematical Sciences, Polytechnic University of Turin\\
    Department of Statistical Methodology, Saryga  \vspace{0.2cm} \\
    Gaëlle Saint-Hilary \\
    Department of Statistical Methodology, Saryga \vspace{0.2cm} \\
    Mauro Gasparini \\
    Department of Mathematical Sciences, Polytechnic University of Turin
    \vspace{0.2cm} \\
    Pavel Mozgunov \\
    MRC Biostatistics Unit,
    University of Cambridge \\
    Department of Statistical Methodology, Saryga}
  \maketitle
} \fi

\if0\blind
{
  \bigskip
  \bigskip
  \bigskip
  \begin{center}
    {\LARGE \bf On the interplay between prior weight and variance of the robustification component in Robust Mixture Prior Bayesian Dynamic Borrowing approach}
\end{center}
  \medskip
} \fi

\begin{abstract}
\noindent
Robust Mixture Prior (RMP) is a popular Bayesian dynamic borrowing method, which combines an informative historical distribution with a less informative component (referred as \textit{robustification component}) in a mixture prior to enhance the efficiency of hybrid-control randomized trials. Current practice typically focuses solely on the selection of the prior weight that governs the relative influence of these two components, often fixing the variance of the robustification component to that of a single observation. In this study we demonstrate that the performance of RMPs critically depends on the \textit{joint} selection of both weight and variance of the robustification component. In particular, we show that a wide range of weight-variance pairs can yield practically identical posterior inferences (in particular regions of the parameter space) and that large variance robust components may be employed without incurring  in the so called \textit{Lindley’s paradox}.
We further show that the use of large variance robustification components leads to improved asymptotic type I error rate control and enhanced robustness of the RMP to the specification of the location parameter of the robustification component.  Finally, we leverage these theoretical results to propose a novel and practical hyper-parameter elicitation routine.

\end{abstract}

\noindent%
{\it Keywords:}  Robust Mixture Prior, Bayesian Dynamic Borrowing, Lindley's paradox, Clinical Trials, Bayesian Methods
\vfill

\newpage

\section{Introduction}
Leveraging historical information in clinical trials is particularly valuable in contexts like rare diseases \supercite{dunoyer_accelerating_2011} and pediatric trials \supercite{dunne_extrapolation_2011-1, schoenfeld_bayesian_2009, Rover}, where recruiting large patient populations is challenging. Bayesian designs are appealing as they allow incorporating available knowledge into prior distributions. However, including external data raises challenges, such as quantifying heterogeneity between external and current data, which can lead to biased estimates and poor operating characteristics if not properly addressed. \vspace{0.2cm} \\ \noindent
Bayesian dynamic borrowing (BDB) sets out to solve such issue by dynamically discounting the use of external information based on a measure of heterogeneity between the prior distribution and the observed data. 
Several borrowing strategies have been proposed over the years such as Power priors \supercite{Hobbs2011, Ibrahim2015}, commensurate priors \supercite{Pocock1976} and Robust Mixture Prior (RMP) \supercite{Kleyner1996, Mutsvari2016}, all of them requiring the specification of a tuning parameter quantifying the amount of borrowing (called {\em knowledge factor} in an early non clinical reference \supercite{Kleyner1996}). A thorough review of the available borrowing methods can be found in Van Rosmalen et al. \supercite{VanRosmalen} and Viele et al. \supercite{Viele2014}.  
%Many strategies have been proposed for a data-dependent selection of the hyper-parameters \supercite{Gravestock2017, Ollier2020, Zhang2023, Yang2023}; however from a conceptual point of view the legitimacy of the latter procedure is arguable, since the parameters are part of \textit{prior} distributions, and should therefore be selected \textit{prior} to observing data. In case the parameters elicitation is data-agnostic,
Among them, Robust Mixture Prior (RMP) \supercite{Schmidli2014, Mutsvari2016}, is acknowledged as one of the most versatile options due to its natural ability of dynamically discounting the amount of borrowed information as the prior-data conflict increases. 
Examples of practical use of RMP in different contexts of application can be found in literature, e.g. bringing adult information to inform treatment effect on a pediatric trial \supercite{Rover}, exploiting expert opinion to inform a prior distribution for a treatment effect \supercite{Mutsvari2016}, borrowing historical information to predict a treatment effect on a primary endpoint based on a surrogate endpoint \supercite{Fougeray2024, Saint-Hilary2019} or borrowing external control data to discount sample size in the control arm \supercite{Roychoudhury2020}. \vspace{0.2 cm} \\  \noindent
The idea behind RMP is to construct a prior distribution for the parameter of interest by combining an informative component, derived from external information, and a \textit{robustification} high-variance component 
%(also known as \textit{dilution} component \supercite{Berger1986})
in a mixture distribution. The advantage of this approach is that the information contained in the informative component of the mixture impacts the posterior inference in a dynamic way, i.e. mostly in case of agreement between historical and current data, while it is progressively disregarded as the prior-data conflict increases \supercite{Schmidli2014}. \vspace{0.2 cm} \\  \noindent
The main object of investigation of this paper are robust mixtures of normal priors, called normal RMPs, which are vastly used in case of normally distributed (or approximately normally distributed) endpoints. 
In particular, we will focus on the case in which the informative component of the RMP is a single normal distribution with known mean and variance, and is combined with a robust normal component with higher variance. 
%An extension of the proposed methodology to the more complex case where the informative component is a mixture of normal distributions is possible, albeit out of the scope of this work. \\ 
In this context, three parameters must be specified, namely \textit{i) weight} of the robustification component of the mixture prior, \textit{ii) location} of the robustification component and \textit{iii) variance} of the robustification component. Although it has been shown that all these three factors impact the operating characteristics (see Weru et al. \supercite{Weru2024}), it is common to focus solely on the selection of the mixture weight related to the informative component (referred to as ``mixture weight''), regulating the amount of information to be borrowed. The latter is commonly pre-specified based on the stakeholder degree of confidence in the historical source, while all the other parameters are commonly fixed. For the variance of the robustification component of the mixture it has been argued that extremely large variances should be avoided \supercite{Mutsvari2016, Weru2024, Callegaro2023}, as they can lead to borrowing of historical information even in case of extreme inconsistency between historical and concurrent data. To avoid this situation, robust weakly informative components have generally been preferred and unit information priors (UIP) \supercite{Schmidli2014} have become a common choice. Using weakly informative robustification components, however, has some drawbacks, in particular \textit{i)} it is sensitive to the choice of the location of the robustification component \supercite{Weru2024}, and \textit{ii)} it causes an inflation of type I error rate in case of the major inconsistency between historical and current data. \vspace{0.2 cm} \\  \noindent\noindent
In this work, we demonstrate that the borrowing properties of the RMP are defined by the \textit{joint} specification of prior weight and variance of the robustification component and these two parameters should be chosen together. 
We theoretically demonstrate that RMP with high-variance robustification components is a viable choice, provided a jointly optimized selection of prior weight and variance of the robustification component.  We argue that this approach is advantageous as \textit{ i)} it practically makes the choice of the location of the robustification component impactless and \textit{ ii)} it effectively prevents from the asymptotic inflation of the type I error rate, which arises - in the case of weakly informative robustification components - when major inconsistency between historical and current data is observed.
\vspace{0.2 cm} \\
\noindent
The manuscript is organized as follows: Sections \ref{2}--\ref{6} focus on the normal setting. Specifically, Section \ref{2} introduces the RMP model and its application in the normal setting; Section \ref{3} presents the motivation for this work; Section \ref{4} details the theoretical findings for the normal setting; Section \ref{5} provides a proof-of-concept analysis highlighting the key benefits of the proposed methodology; and Section \ref{6} outlines a novel procedure for hyper-parameter selection. Section \ref{7} discusses the extension to the binary case with the Beta RMP, while Section \ref{8} presents the extension to scenarios in which the informative component of the RMP is itself a mixture. Finally, Section \ref{9} concludes with a discussion.

\section{Methodology}
\label{2}

\subsection{Setting}
\label{2.1}

\subsubsection{Bayesian Design of a Randomized Controlled Trial (RCT)}
\label{2.1.1}

Consider a randomized controlled trial (RCT) evaluating a novel treatment against placebo or standard of care. Let $X_t$ and $X_c$ denote the normally distributed mean treatment and control responses with unknown means $\theta_t$ and $\theta_c$, and known variances $\sigma_t^2 = s^2/n_t$ and $\sigma_c^2 = s^2/n_c$, where $s$ is the common variance of individual responses and $n_j$ $(j=t,c)$ the arm-specific sample sizes.  
\vspace{0.2cm} \\
\noindent
The treatment effect $\delta = \theta_t - \theta_c$ is the parameter of interest, with $H_0: \delta = 0$ tested against $H_A: \delta > 0$. Priors $\pi_t(\cdot)$ and $\pi_c(\cdot)$ are specified for $\theta_t$ and $\theta_c$.  
\vspace{0.2cm} \\
\noindent
Trial success is declared when the posterior probability of a positive treatment effect exceeds a prespecified threshold:
\begin{equation}
\label{success RCT}
\mathbb{P}_{\pi_c, \pi_t}\big( \delta > 0 \;|\; x_c, x_t \big) > 1 - \eta,
\end{equation}
where $x_c$ and $x_t$ are observed mean responses. The threshold $1 - \eta$ represents the required posterior evidence for efficacy; with smaller $\eta$ values imply more stringent criteria.

\subsubsection{Frequentist and Bayesian Operating Characteristics}
\label{2.1.2}

The \textit{type I error rate}, the probability of rejecting $H_0$ when $\delta=0$, is computed by integrating the success condition over the data likelihoods:
\begin{equation}
\label{t1e}
\alpha(H) = \iint_{\mathbb{R}^2} 
\vmathbb{1} \Big\{\mathbb{P}_{\pi_c, \pi_t}(\delta>0|x_c,x_t) > 1-\eta\Big\}
f_{X_c}(x_c| \theta_c=H) f_{X_t}(x_t| \theta_t = H) \, dx_c \, dx_t,
\end{equation}
where $\vmathbb{1} (\cdot)$ is the indicator function, and $f_{X_c}$, $f_{X_t}$ denote the sampling distributions.  
\textit{Power} is obtained analogously under $\theta_t = H + \delta^*$ and $\theta_c = H$, for a target effect $\delta^* > 0$.  
\vspace{0.2cm} \\
\noindent
type I error rate and power are frequentist quantities, as they condition on fixed parameter values. To assess Bayesian designs more comprehensively, Best et al. \supercite{Best2023} proposed averaging $\alpha$ over a \textit{design prior} $\Pi_c$, namely:

\begin{equation}
\label{avg.t1e}
\alpha^{\Pi_c}_{\text{avg}} = \int_{\mathbb{R}} \alpha(t)\,\Pi_c(t)\,dt.
\end{equation}

\noindent
A \textit{design prior} is the prior distribution used during the \emph{planning} of the trial to reflect plausible values for the parameters, which allows evaluation of Bayesian operating characteristics such as average Type~I error and power. It is not necessarily the same as the prior used in the \emph{analysis} of the trial, which represents the formal beliefs applied to the data once observed. The design prior is primarily a tool for trial design and simulation, whereas the analysis prior is used for inference and decision-making.

\subsubsection{Posterior Estimation Metrics}
\label{2.1.3}

Besides testing, performance is evaluated through estimation metrics. The posterior median $\hat{\delta}$ serves as point estimate, and bias, variance, and mean squared error (MSE) quantify its accuracy (see Supplementary Material for formulas).

\subsection{Robust Mixture Prior (RMP)}
\label{2.2}

Let $\pi_{\text{inf}}(\cdot)$ be an informative prior for $\theta_c$. The \textit{Robust Mixture Prior (RMP)} combines this with a weakly informative or non-informative robustification component $\pi_{\text{rob}}(\cdot)$:
\begin{equation}
\label{RMP}
\pi_c(\theta_c) = \omega\,\pi_{\text{inf}}(\theta_c) + (1-\omega)\,\pi_{\text{rob}}(\theta_c),
\end{equation}
where $\omega \in [0,1]$ is the prior weight on the informative component. The robustification term downweights historical information when inconsistent with current data.  
\vspace{0.2cm} \\
\noindent
After observing $x_c$, the posterior is again a mixture:
\begin{equation}
\label{RMP update}
g(\theta_c|x_c) = \tilde{\omega}\,g_{\text{inf}}(\theta_c|x_c) + (1-\tilde{\omega})\,g_{\text{rob}}(\theta_c|x_c),
\end{equation}
where each component posterior is  
$g_\star(\theta_c|x_c) = f(x_c|\theta_c)\pi_\star(\theta_c)/f(x_c|\pi_\star)$,  
with $\star\in\{\text{inf},\text{rob}\}$.  
The updated weight depends on $x_c$ via the formula
\begin{equation}
\label{Weights Update}
\tilde{\omega}(x_c) = 
\frac{\omega f(x_c|\pi_{\text{inf}})}{\omega f(x_c|\pi_{\text{inf}}) + (1-\omega) f(x_c|\pi_{\text{rob}})}.
\end{equation}
A proof of Equation \eqref{RMP update}
and \eqref{Weights Update} is in the Supplementary Material.
\vspace{0.2cm} \\
Equation \eqref{Weights Update} can be expressed equivalently in terms of odds as
\begin{equation}
\label{Odds update}
\tilde{\Omega}(x_c) = \Omega \frac{f(x_c|\pi_{\text{inf}})}{f(x_c|\pi_{\text{rob}})},
\end{equation}
with $\Omega = \omega/(1-\omega)$ and $\tilde{\Omega} = \tilde{\omega}/(1-\tilde{\omega})$. It can be noticed that weights (and odds) adjust borrowing dynamically according to the data’s compatibility with prior information, namely increases when the observed response $x_c$ is compatible with the informative component of the mixture while decreases otherwise.
\vspace{0.2cm} \\
\noindent
Note that in Equations (\ref{Weights Update}) and (\ref{Odds update}), posterior weights and posterior odds are well-defined functions of the observed mean response, conditional on the specified RMP for $\theta_c$. For simplicity, this dependence will be implicitly understood in subsequent sections and explicitly stated only when necessary.

\subsection{Normal Robust Mixture Prior}
\label{2.3}

When both mixture components are Normal,
\[
\pi_{\text{inf}}(\theta_c) = \mathcal{N}(\mu_{\text{inf}}, \sigma^2_{\text{inf}}), \quad
\pi_{\text{rob}}(\theta_c) = \mathcal{N}(\mu_{\text{rob}}, \sigma^2_{\text{rob}} = s^2/n_0),
\]
the conjugacy ensures that the posterior remains a Normal mixture with updated parameters.  
\noindent
Moreover, the corresponding prior predictive distributions are also Normal:
\begin{equation}
\label{predictive}
f(x_c|\pi_\star) = 
\frac{1}{\sqrt{2\pi v_\star^2}} 
\exp\!\left[-\frac{(x_c-\mu_\star)^2}{2v_\star^2}\right],
\quad v_\star^2 = \sigma_\star^2 + \sigma_c^2,
\end{equation}
for $\star \in \{\text{inf}, \text{rob}\}$.  As a consequence, letting $R = v_{\text{rob}}/v_{\text{inf}}$, then Equation (\ref{Odds update}) becomes
\begin{equation}
\label{update normal}
\tilde{\Omega}(x_c) = \beta(\omega, \sigma^2_{\text{rob}})
\exp\!\left\{
- \frac{d^2}{2v_{\text{inf}}^2} + 
\frac{(x_c-\mu_{\text{rob}})^2}{2R^2 v_{\text{inf}}^2}
\right\}.
\end{equation}
In the latter, $\beta\left(\omega, \sigma^2_{\text{rob}}\right)=\Omega/R$, while $d$ represents the realization of the random variable $X_c-\mu_{\text{inf}} \sim \mathcal{N}\left(D,\sigma^2_c\right)$, with mean $D$ representing the true \textit{drift} parameter (also referred to as \textit{prior-data conflict} hereinafter), indicating the level of inconsistency between concurrent data and historical information provided in the informative component of the RMP. Note that defining the function $\beta\left(\cdot \right)$ will become useful in Section \ref{4.3}. \vspace{0.2 cm} \\  \noindent
Equation (\ref{update normal}) shows that the posterior odds $\tilde{\Omega}$ depend on the choice of $\Omega$ (which is a deterministic function of the prior weight $\omega$),  the location parameter of the robustification component $\mu_{\text{rob}}$ and the variance of the robustification component $\sigma_{\text{rob}}^2$.
\vspace{0.2cm} \\
\noindent
Notice that, since the robustification component must be less informative than the informative one, $R>1$ (often $R\gg1$ when $\pi_{\text{rob}}$ is nearly non-informative).

\section{Motivation for the Work}
\label{3}

\subsection{Background}
\label{3.1}

Robust Mixture Priors (RMPs) are widely applied in randomized controlled trials (RCTs) to borrow information for the control arm \supercite{Best2023, Roychoudhury2020, Callegaro2023.bis}. Several approaches exist for specifying the mixture weight $\omega$ \supercite{Zhang2023, Yang2023}, yet the selection of hyperparameters for the robustification component has received limited attention.  
\vspace{0.2cm} \\
\noindent
Large variances for the robustification prior are often adopted to represent minimal prior knowledge; however, such weakly informative choices may retain excessive influence of the informative component even under strong prior–data conflict—an effect known as \textit{Lindley’s paradox} \supercite{Mutsvari2016, Weru2024, Callegaro2023}. Schmidli et al. \supercite{Schmidli2014} proposed mitigating this through a \textit{unit-information prior} (UIP),  namely a distribution which effective sample size (ESS)\supercite{Morita2008} is equal to 1.  
\vspace{0.2cm} \\
\noindent
While practical and commonly used, this approach introduces two main challenges:  
\textit{(i)} the pre-specification of the robustification mean $\mu_{\text{rob}}$, which strongly affects posterior inference \supercite{Weru2024}; and  
\textit{(ii)} the asymptotic inflation of the Type~I error in the presence of substantial discrepancies between the historical and current control data\supercite{Weru2024, Best2023}. 
Here, the term \textit{asymptotic inflation} refers to the progressive increase in the Type~I error rate as the drift parameter $D$ increases, such that the Type~I error approaches~1 as $D \to +\infty$.
\vspace{0.2cm} \\
\noindent
The following case study illustrates these issues in Normal RMPs within hybrid-control RCTs, providing the basis for the theoretical developments in Section \ref{4}.

\subsection{Illustration in a Hypothetical Trial}
\label{3.2}

Consider a two-arm RCT comparing treatment and control (placebo or standard of care). Individual outcomes in both arms follow normal distributions with unit variance ($s=1$), as a consequence the mean responses in the two arms are:
\[
X_t \sim \mathcal{N}(\theta_t, n_t^{-1}), \quad X_c \sim \mathcal{N}(\theta_c, n_c^{-1}).
\]
The trial allocates $n_t = 150$ patients to treatment and $n_c = 50$ to control (3:1 ratio). Trial success is defined by Equation~(\ref{success RCT}) with $\eta = 0.05$.  
\vspace{0.2cm} \\
\noindent
No prior information is available for $\theta_t$, so a non-informative prior $\theta_t \sim \mathcal{N}(\mu_{\text{rob}}, n_0^{-1})$ is used. For $\theta_c$, an informative prior $\mathcal{N}(\mu_{\text{inf}}, n_{\text{inf}}^{-1})$ with effective sample size $n_{\text{inf}} = 100$ and mean $\mu_{\text{inf}} = 0$ is combined with a non-informative prior $\mathcal{N}(\mu_{\text{rob}}, n_0^{-1})$ through an RMP with weight $\omega$.  
\vspace{0.2cm} \\
\noindent
Performance metrics include the type I error rate (Equation~\ref{t1e}), power (for target $\delta^* = 0.31$), and the average posterior weight $\tilde{\omega}$, obtained by integrating Equation \eqref{Weights Update} over the data likelihood.
\vspace{0.2cm} \\
\noindent
Different RMP configurations are examined, considering mixture weights $\omega \in \{0.5, 0.9\}$ to represent, respectively, moderate and strong confidence in the historical information. Six sub-scenarios are defined by varying the hyperparameters of the robustification component. Specifically, the location parameter is set to $\mu_{\text{rob}} \in \{-2, 0, 2\}$, while the variance takes values $\sigma^2_{\text{rob}} \in \{1, 10^{100}\}$, the former corresponding to a unit-information prior and the latter approximating an improper prior. A reference setting with $\omega = 0$ and $\sigma^2_{\text{rob}} = 10^{100}$ represents a standard non-informative Bayesian design. Performance metrics are assessed across a range of \textit{drift} values $D$.

\subsection{Analysis}
\label{3.4}

\begin{figure}[H]
	\begin{subfigure}[type I error rate with $\omega=0.5$]{
		\centering
		\includegraphics[width=0.45\textwidth]{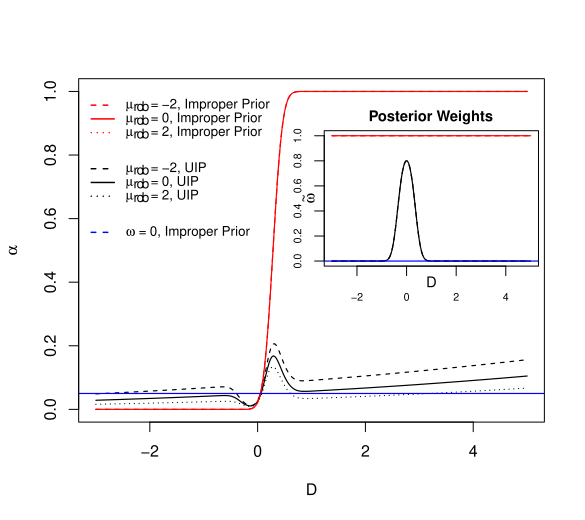}
        \label{w=0.5.fig}}
	\end{subfigure} \hfill
	\begin{subfigure}[type I error rate with $\omega=0.9$]{
		\centering
		\includegraphics[width=0.45\textwidth]{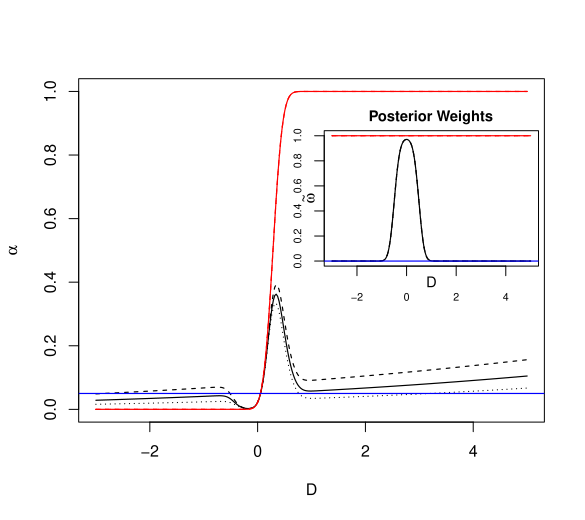}
        \label{w=0.9.fig}}
	\end{subfigure}
	\caption{type I error rate $\alpha(D)$ under different RMP parameterizations. Red curves: improper priors ($\sigma^2_{\text{rob}} = 10^{100}$). Black curves: unit-information priors ($\sigma^2_{\text{rob}} = 1$). Line styles denote values of $\mu_{\text{rob}}$. Panel (\ref{w=0.5.fig}): $\omega=0.5$; Panel (\ref{w=0.9.fig}): $\omega=0.9$.}
    \label{motivating.exe.fig}
\end{figure}

Figure~\ref{motivating.exe.fig} displays the type I error rate as a function of the drift parameter $D$ for $\omega=0.5$ (left) and $\omega=0.9$ (right), under varying $\mu_{\text{rob}}$ and $\sigma^2_{\text{rob}}$.

\noindent
When a UIP is used as robustification component, type I error rate decreases near $D \approx 0$, reflecting improved borrowing when historical and current data agree ($\tilde{\omega}\gg0$). As $|D|$ increases, borrowing diminishes; however, intermediate drifts can still yield residual borrowing ($\tilde{\omega}>0$), biasing control estimates and inflating type I error rate for positive drifts or deflating it for negative ones.  
\vspace{0.2cm} \\
\noindent
Under extreme prior–data conflict ($|D|$ large), borrowing vanishes ($\tilde{\omega}\approx0$), yet instead of stabilizing near the nominal level, type I error rate asymptotically diverges toward 1 for $D\to+\infty$ and 0 for $D\to-\infty$. This counterintuitive behavior motivates further theoretical investigation.  
\vspace{0.2cm} \\
\noindent
Figure~\ref{motivating.exe.fig} also shows that, although all UIP-based RMPs share similar asymptotic trends, the choice of $\mu_{\text{rob}}$ systematically shifts type I error rate: larger $\mu_{\text{rob}}$ increases it uniformly across $D$, while smaller values decrease it. This sensitivity to $\mu_{\text{rob}}$ forms a second point of interest.  
\vspace{0.2cm} \\
\noindent
When the robustification component is nearly improper ($\sigma^2_{\text{rob}} = 10^{100}$), borrowing persists regardless of how strong the prior–data conflict  is ($\tilde{\omega} = 1$ across all the D-space), illustrating \textit{Lindley’s paradox} \supercite{Mutsvari2016, Weru2024, Callegaro2023}. Here, type I error rate remains near 0 for $D < -0.2$, increases sharply to 1 for $-0.2 \le D \le 0.5$, and stays at this level thereafter, with negligible dependence on $\mu_{\text{rob}}$.

\subsection{Research Questions}
\label{3.5}

n section \ref{3.4} we have shown that there are some issues related to the use RMP in the context of hybrid control RCT. These are: 

\begin{enumerate}
    \item The asymptotic inflation of type I error for large positive values of prior-data conflict, when weakly informative robustification components are employed.
    \item The sensitivity of the operating characteristics to the choice of $\mu_{\text{rob}}$, when weakly informative robustification components are employed.
    \item The apparent failure in discounting information borrowing as the prior-data conflict increases, when large variance robustification components are used (Lindley's paradox).
\end{enumerate}

\noindent
In the next sections the cause of these issues will be theoretically investigated, and a solution to all of them will be proposed.

\section{Analytical results}
\label{4}

\subsection{Asymptotic inflation of type I error rate}
\label{4.1}

The cause of the asymptotic type I error rate inflation, along with the conditions under which the latter is prevented are investigated in Theorem \ref{theo.t1e}. In particular, it is proven that type I error rate inflation occurs when an upwards bias is induced by the robustification component $\pi_\text{rob}$ of the RMP on the posterior mean for the treatment difference. 
For a fixed value of the mixture weight $\omega$, this bias is inversely proportional to the variance of the robustification component $\sigma^2_{\text{rob}}$, and in particular it is null if the latter diverges to $+\infty$ at least as fast as the drift parameter $D$. Under this condition, an asymptotic control of the type I error rate is achieved, thus making the choice of large variance robustification components in RMPs particularly attractive.

\begin{theorem} \label{theo.t1e} Consider a RCT where mean control and treatment responses are normal $X_c \sim \mathcal{N} \left(\theta_c, \sigma^2_c \right)$, $X_t \sim \mathcal{N} \left(\theta_t, \sigma^2_t \right)$, and assume $\sigma^2_t = K\sigma^2_c$ (where $K^{-1}$ is the randomization ratio, assumed > 1). Assume a RMP  $\pi_c(\theta_c) = \omega \pi_\text{inf}(\theta_c)+(1-\omega) \pi_\text{rob}(\theta_c)$ is used for the control parameter, where $\pi_\text{inf}(\theta_c)$ and $\pi_\text{rob}(\theta_c)$ are the PDF of normally distributed random variables with parameters $\mu_{\text{inf}}$, $\sigma^2_{\text{inf}}$ and $\mu_{\text{rob}}$, $\sigma^2_{\text{rob}}$ respectively; while a normal prior distribution $\theta_t \sim \mathcal{N} \left(\mu_t, \sigma^2_\text{rob} \right)$ is given to the treatment parameter. Consider the type I error rate $\alpha \left( \cdot \right)$ as defined in Equation (\ref{t1e}), corresponding to the null hypothesis $H_0: \theta_c = \theta_t = D + \mu_\text{inf}$, where $D=\theta_c-\mu_{\text{inf}}$ is the drift parameter.
Then the following hold:
\begin{equation*}
    \lim_{D \rightarrow + \infty} \alpha \left( D + \mu_{\text{inf}}\right) = \eta \; \; \; \Longleftrightarrow \; \; \;  \lim_{D \rightarrow +\infty} \frac{D}{\sigma^2_{\text{rob}}} = 0
\end{equation*}
\end{theorem}

\noindent
A formal proof of Theorem \ref{theo.loc} can be found in the supplementary material. A numerical validation of this result is shown in Section \ref{5}, while a practical use of the latter in parameter selection can be found in Section \ref{6}.

\subsection{The impact of the selection of \textbf{$\mu_{\text{rob}}$}}
\label{4.2}

The robustification component of the mixture acts to \textit{robustly} model the tails of the informative component's prior distribution.  Ideally, it represents a lack of prior knowledge, thereby hindering precise elicitation of its location parameter $\mu_{\text{rob}}$. This choice, however, may significantly impact the posterior inference, as demonstrated by Weru et al. \supercite{Weru2024}. \\
Theorem \ref{theo.loc} investigates the condition under which the choice of $\mu_{\text{rob}}$ becomes impact-less in the posterior inference, showing that employing robustification components with large variances effectively prevents from bias stemming from the chosen location, enabling then the use of any convenient value for $\mu_{\text{rob}}$.

\begin{theorem} \label{theo.loc} Consider a normal random variable modeling the mean control response $X_c \sim \mathcal{N} \left(\theta_c, \sigma^2_c \right)$, and assume two distinct RMPs are used for the underlying parameter $\theta_c$, namely
\begin{equation*}
    \pi^{(1)}_c(\theta_c) = \omega \pi_\text{inf}(\theta_c)+(1-\omega) \pi_\text{rob}^{(1)}(\theta_c) \; \; \; \; \; \; \; \; \pi^{(2)}_c(\theta_c) = \omega \pi_\text{inf}(\theta_c)+(1-\omega) \pi_\text{rob}^{(2)}(\theta_c)
\end{equation*} 
where $\pi_\text{inf}(\theta_c)$ and $\pi^{(i)}_\text{rob}(\theta_c)$ are the PDF of normally distributed random variables with parameters $\mu_{\text{inf}}$, $\sigma^2_{\text{inf}}$ and $\mu^{(i)}_{\text{rob}}$, $\sigma^2_{\text{rob}}$ respectively with $i \in \{ 1,2 \}$. \\
Consider the posterior distributions $ g(\theta_c  |  x_c, \pi^{(1)}_c)$ and $ g(\theta_c  |  x_c, \pi^{(2)}_c)$, then
\begin{equation*}
    \lim_{\sigma^2_{\text{rob}} \rightarrow + \infty}  g(\theta_c  |  x_c, \pi^{(1)}_c) = \lim_{\sigma^2_{\text{rob}} \rightarrow + \infty} g(\theta_c  |  x_c, \pi^{(2)}_c) \; \; \; \; \; \; \; \; \; \; \; \; \; \; \forall x_c \in \mathbb{R}
\end{equation*}
\end{theorem}

\noindent
A formal proof of Theorem \ref{theo.loc} can be found in the supplementary material. A numerical validation of this result is presented in Section \ref{5}, while a practical use of the latter in parameter selection is proposed in Section \ref{6}.

\subsection{The Lindley's paradox}
\label{4.3}

The phenomenon termed ``Lindley's paradox'' within the context of robust mixture priors (RMPs) describes the counterintuitive situation where full borrowing (defined as $\tilde{\omega}=1$) occurs despite significant prior-data conflict.  Literature suggests this arises when the RMP's robustification component is improper \supercite{Mutsvari2016, Weru2024, Callegaro2023}.  This occurs because the prior predictive distribution for the robustification component, shown in Equation (\ref{predictive}), becomes improper ($R \rightarrow +\infty$), leading to $\Tilde{\omega}=1$ for all observed control responses $x_c$ according to Equation (\ref{update normal}). In Theorem \ref{theo.lindley} we show that this behavior is due to the hidden underlying assumption that the mixture weight $\omega$ is fixed and independent on the choice of $\sigma^2_{\text{rob}}$. We find that relaxing this assumption, effectively prevents from the occurring of Lindley's paradox. 
\begin{theorem} \label{theo.lindley}
    Consider a normal random variable $X_c \sim \mathcal{N} \left(\theta_c, \sigma^2_c \right)$, and assume a RMP is used for the parameter $\theta_c$, namely $\pi_c(\theta_c) = \omega \pi_\text{inf}(\theta_c)+(1-\omega) \pi_\text{rob}(\theta_c)$, where $\pi_\text{inf}(\theta_c)$ and $\pi_\text{rob}(\theta_c)$ are the PDF of normally distributed random variables with parameters $\mu_{\text{inf}}$, $\sigma^2_{\text{inf}}$ and $\mu_{\text{rob}}$, $\sigma^2_{\text{rob}}$ respectively. The following hold:
    \begin{enumerate}
        \item if $\Omega < + \infty$, then
        \begin{equation*}
        \lim_{\sigma^2_{\text{rob}} \rightarrow + \infty} \Tilde{\omega} \left(x_c, \pi_\text{inf}(\theta_c), \pi_\text{rob}(\theta_c), \omega \right) = 1 \; \; \; \; \; \; \; \forall x_c \in \left( -\infty, +\infty \right)
        \end{equation*}
        
        \item if $\Omega \sim O(R)$ for $\sigma^2_{\text{rob}} \rightarrow + \infty$, then
        \begin{equation*}
            \lim_{\sigma^2_{\text{rob}} \rightarrow + \infty} \Tilde{\omega} \left(x_c, \pi_\text{inf}(\theta_c), \pi_\text{rob}(\theta_c), \omega \right) \neq 1 \; \; \; \; \; \; \; \forall x_c \in \left( -\infty, +\infty \right)
        \end{equation*}    
    \end{enumerate}
\end{theorem}
\noindent
The preceding theorem demonstrates that Lindley's paradox arises, as $\sigma^2_{\text{rob}} \rightarrow +\infty$, when the prior weight $\omega$ (or prior odds $\Omega$) is fixed independently of $\sigma^2_{\text{rob}}$. Conversely, if $\omega$ and $\sigma^2_{\text{rob}}$ are jointly selected such that the prior odds $\Omega$ are of the same order of magnitude as $R$ - as $\sigma^2_{\text{rob}} \rightarrow +\infty$ - then Lindley's paradox is avoided. 
The latter holds because as $\sigma^2_{\text{rob}} \to \infty$, the posterior odds $\tilde{\Omega}$ can be written following Equation (\ref{eq:T2.1}) as
\begin{equation} \label{odds_asy}
\tilde{\Omega}(x_c; \omega, \sigma^2_{\text{rob}}) = \beta(\omega, \sigma^2_{\text{rob}}) \times \exp\left[-\frac{d^2}{2v^2_{\text{inf}}}\right] \;,
\end{equation}
where the influence of the RMP on the posterior odds is entirely captured by the function $\beta(\omega, \sigma^2_{\text{rob}})$. 
As a consequence, all combinations of $\omega$ and $\sigma^2_{\text{rob}}$ yielding $\beta(\omega, \sigma^2_{\text{rob}}) = \beta^*$ share the same ``borrowing profile'', resulting in identical posterior odds (and thus, posterior weights) for any observed value $x_c$. \vspace{0.2 cm} \\  \noindent\noindent
\noindent
The parameter $\beta^*$ governs the RMP's flexibility in borrowing information across the $x_c$ space, determining the rate at which posterior weights decrease with increasing prior-data conflict.  Specifically, it represents the posterior odds when no drift is observed, quantifying the maximum borrowing achievable by the RMP.  Therefore, $\beta^*$ will be referred to as the \textit{borrowing strength}. 
It is important to note that while these pairs yield identical posterior weights, posterior inference for $\theta_c$ could differ in principle across RMPs due to variations in $g_{\text{rob}}(\theta_c | x_c, \pi_{\text{rob}})$, resulting from differing choices of $\mu_{\text{rob}}$ and $\sigma^2_{\text{rob}}$. However, as $\sigma^2_{\text{rob}} \to \infty$, the robust posterior becomes independent of $\mu_{\text{rob}}$, leading to similar inference for $\theta_c$ across all pairs across the entire control response parameter space. \vspace{0.2 cm} \\  \noindent\noindent
\noindent
Note that the asymptotic approximation of posterior odds in Equation (\ref{odds_asy}) is valid only when $R \gg 1$ ($v_{\text{rob}} \gg v_{\text{inf}}$), a reasonable assumption given the robustification component of the RMP is specifically designed for robustification.

\section{Practical considerations}
\label{5}

Using the same trial design considered in Section \ref{3.2}, in the following sections we will focus on the validation of the use of the RMPs with large variance robustification components in the context of unbalanced RCT with hybrid control arms. 

\subsection{Overcoming Lindley's paradox} \label{5.1}
\begin{figure}[H]
   \centering
   \includegraphics[width=0.65\linewidth]{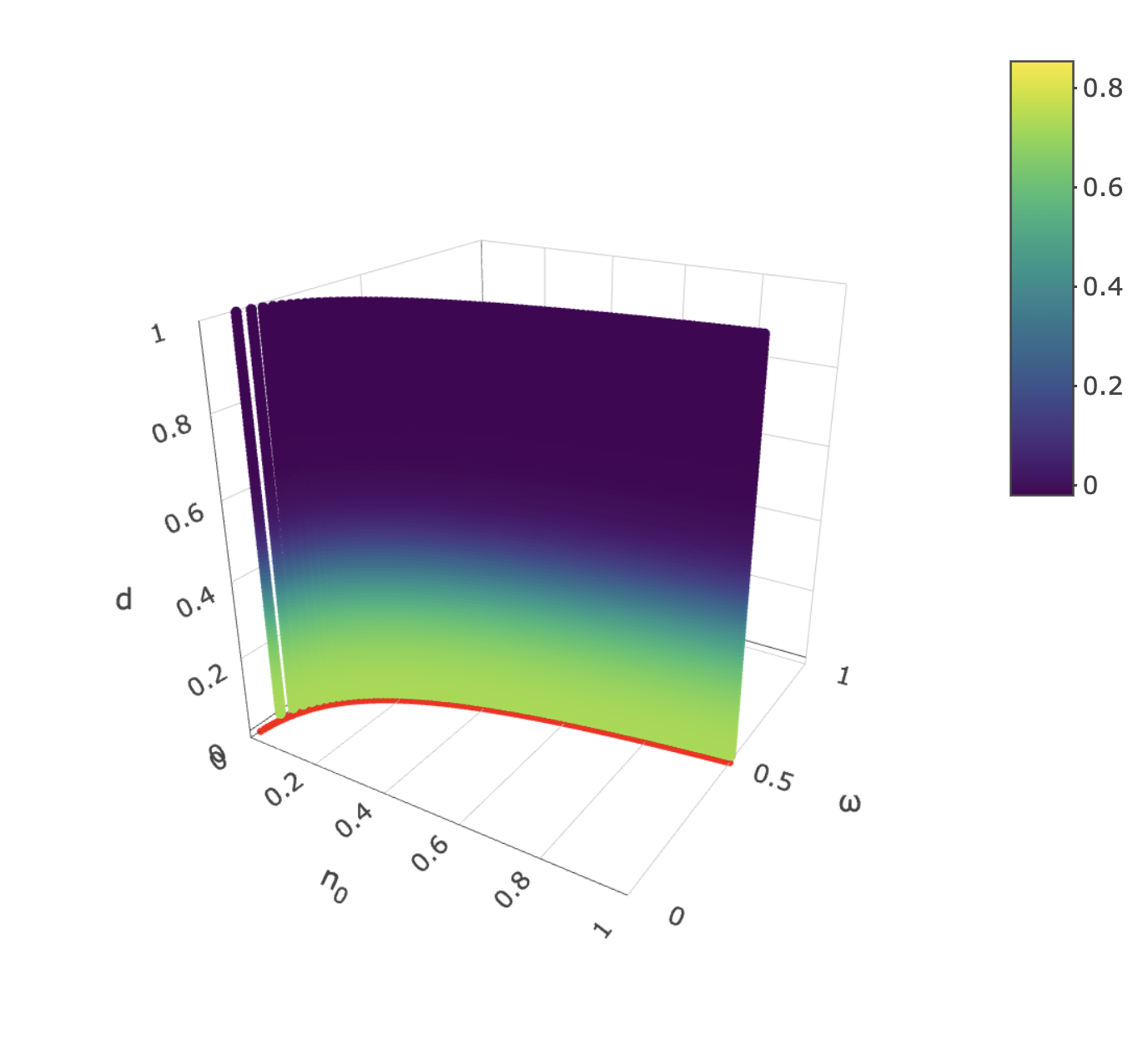}
   \caption{Posterior weight $\Tilde{\omega}$ as a function of effecive sample size of the robust component $n_0$, prior weight $\omega$ and observed control response $x_c$. The red curve in the $(n_0, \omega)$ represents all RMPs with $\beta^{*}=5.83$.}
   \label{3d}
\end{figure}

In Section \ref{4.3} it has been proven that different pairs $(\omega, \sigma^2_{\text{rob}})$ may induce the same posterior weights distribution on the control response space. The latter is illustrated in Figure \ref{3d}.

%\begin{comment}
%\end{comment}

\noindent
Figure \ref{3d} presents a three-dimensional representation with parameters $\omega$ and $n_0=\sigma^{-2}_{\text{rob}}$ on the horizontal axes and the observed control response $x_c$ on the vertical axis.  The red curve embedded in the $(\omega, n_0)$ plane delineates the set of parameter pairs $(\omega, n_0)$ satisfying $\beta(\omega, n_0) = 5.83$, each representing a distinct RMP. Notice that this value has been specifically selected so to include the pair $\omega=0.5$, $n_0=1$, so that
\begin{equation}\label{condition_beta_star}
    \beta^* = \beta(0.5, 1)=  \frac{\frac{0.5}{1-0.5}}{\sqrt{\frac{1+1/50}{1/100+1/50}}}.
\end{equation}

\noindent
The figure was generated by varying the effective sample size of the robust component over the interval $(0.01, 1)$ with a step of $0.01$. For each value, the prior weight $\omega$ was determined to satisfy Equation~\ref{condition_beta_star}, and the posterior odds were computed for each pair $(\omega, n_0)$ using Equation~\ref{Odds update}. The posterior weights were then obtained using the formula $ \Omega = 1/(1+\Omega)$. The vertical colored lines in the figure depict the posterior weights $\tilde{\omega}$ as a function of $x_c$ for all RMPs considered along the red curve, the yellow color indicating a posterior weight of 1 (full borrowing) and the blue color indicating a posterior weight of 0 (no borrowing).
\vspace{0.2cm} \\
\noindent
The vertical lines originating from each point on the red curve exhibit a continuous color gradient along the $x_c$ axis, indicating that the posterior weights $\tilde{\omega}$, as a function of the control response $x_c$, depend solely on the chosen value of $\beta^{*}$. Consequently, all pairs $(\omega, n_0)$ yielding the same $\beta^{*}$ correspond to identical posterior weight profiles. \vspace{0.2cm}
\vspace{0.2cm} \\
\noindent
These observations suggest that Lindley's paradox is effectively mitigated by a joint selection of $\omega$ and $\sigma^2_{\text{rob}}$. Specifically, the posterior weight profile characteristic of any RMP with a weakly informative robustification component (e.g, UIP) can be replicated using robustification components with arbitrarily large variance. Further visualizations of posterior weights under varying $\beta^{*}$ values are provided in the supplementary materials.

\subsection{Overcoming asymptotic type I error rate inflation}
\label{5.2}

While the preceding analysis demonstrates that a set of RMPs share a common posterior weight profile $\tilde{\omega}$, this does not guarantee identical posterior inferences on the control parameter $\theta_c$. Posterior inference is influenced not only by posterior weights but also by the posterior distributions of the individual RMP components, which are functions of their hyper-parameters. \vspace{0.2cm}\\
\noindent
In this section, an analysis of the frequentist operating characteristics is conducted, with specific attention to the problem of asymptotic type I error rate inflation. In addition, the link between the latter and the posterior inference metrics (bias, variance and MSE) is discussed. 

%\begin{comment}
    \begin{figure}[H]
	\begin{subfigure}[type I error rate]{
		\centering
		\includegraphics[width=0.45\textwidth]{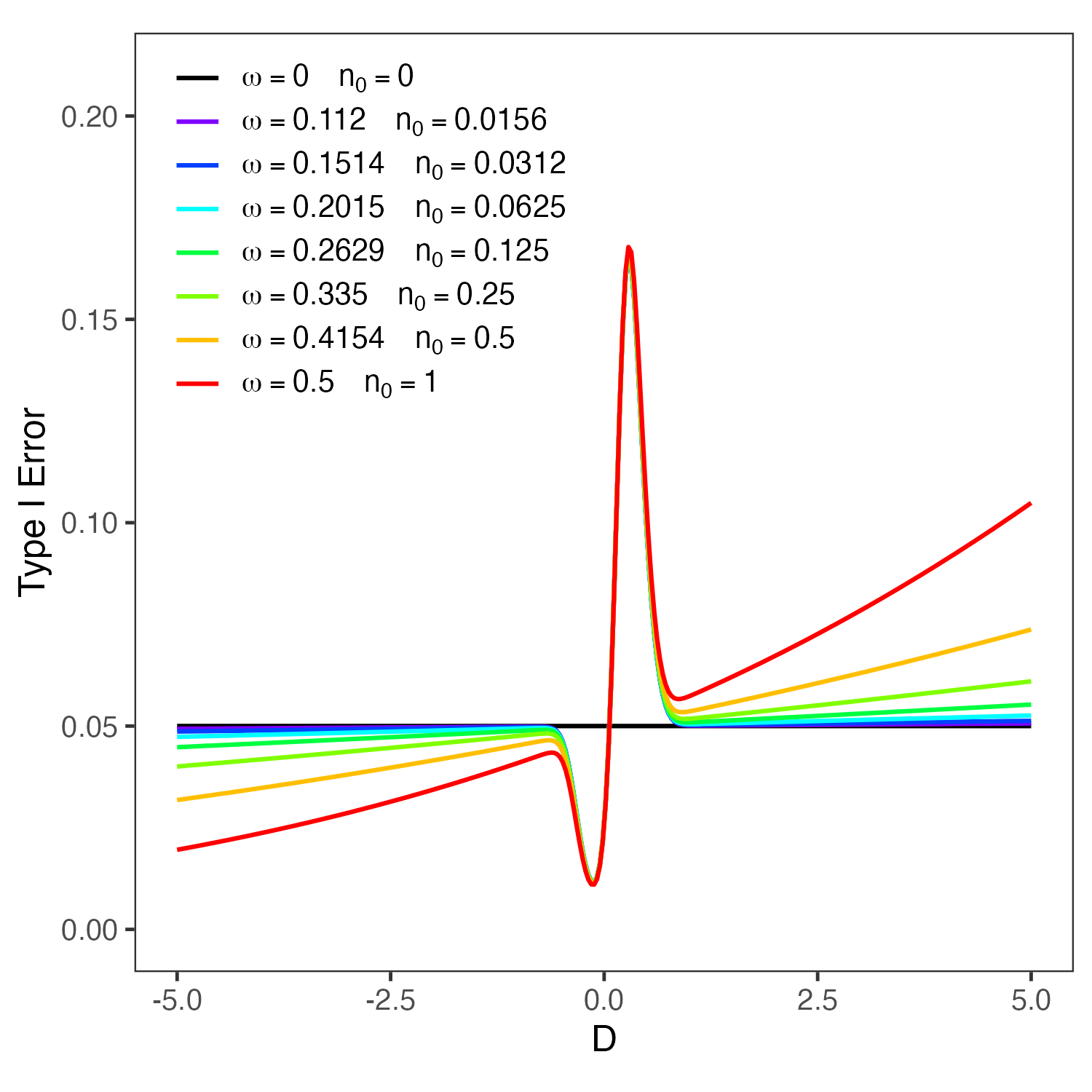}
        \label{t1e.fig}}
	\end{subfigure} \hfill
	\begin{subfigure}[Power]{
		\centering
		\includegraphics[width=0.45\textwidth]{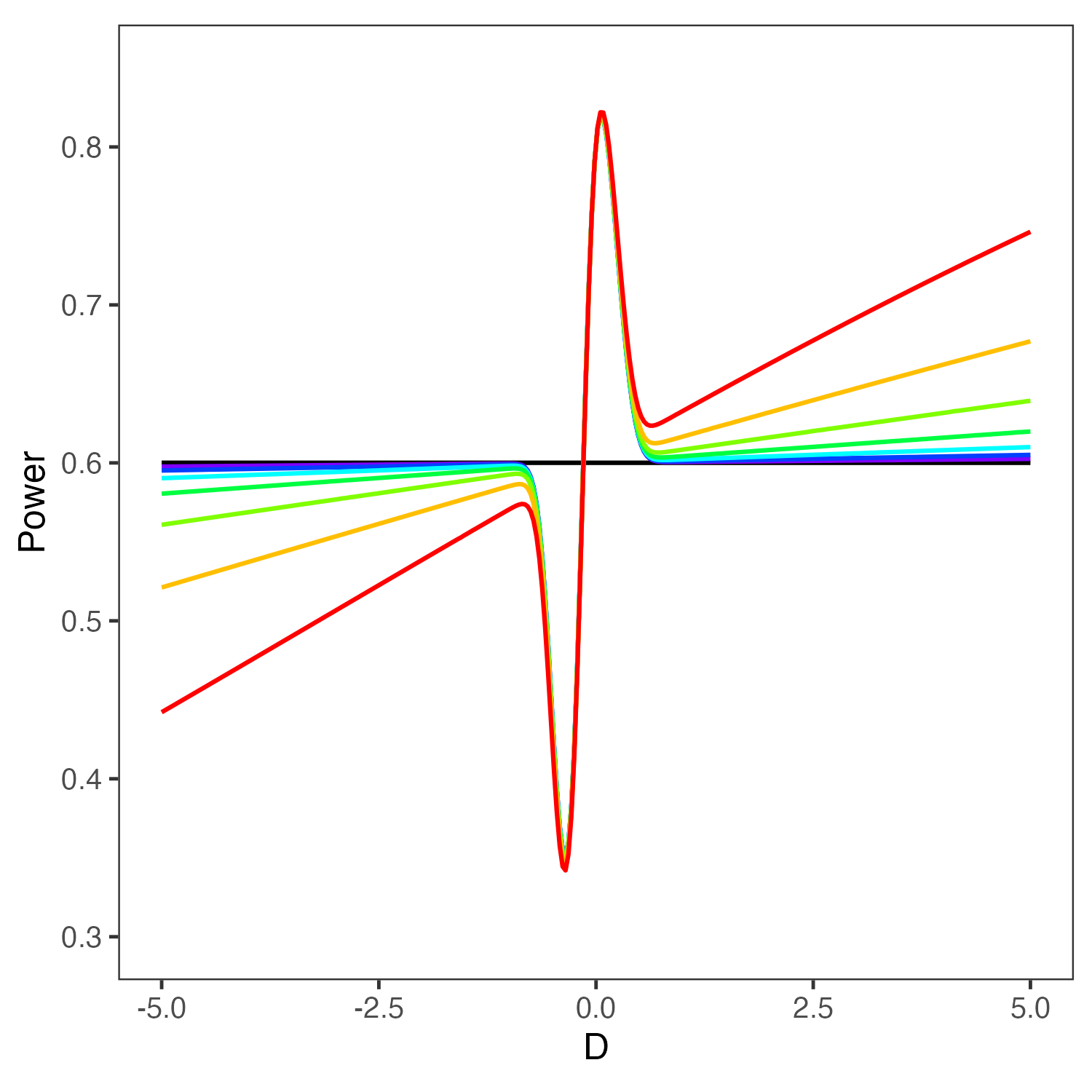}
        \label{pow.fig}}
	\end{subfigure}
	\caption{Panel (a): type I error rate. Panel (b): power under $\delta^{*}=0.31$. Colors represent different couples of $(\omega, n_0)$, corresponding to $\beta=5.83$.}
    \label{OC.fig}
\end{figure}
%\end{comment}

\begin{figure}[H]
	\begin{subfigure}[Bias]{
		\centering
		\includegraphics[width=0.3\textwidth]{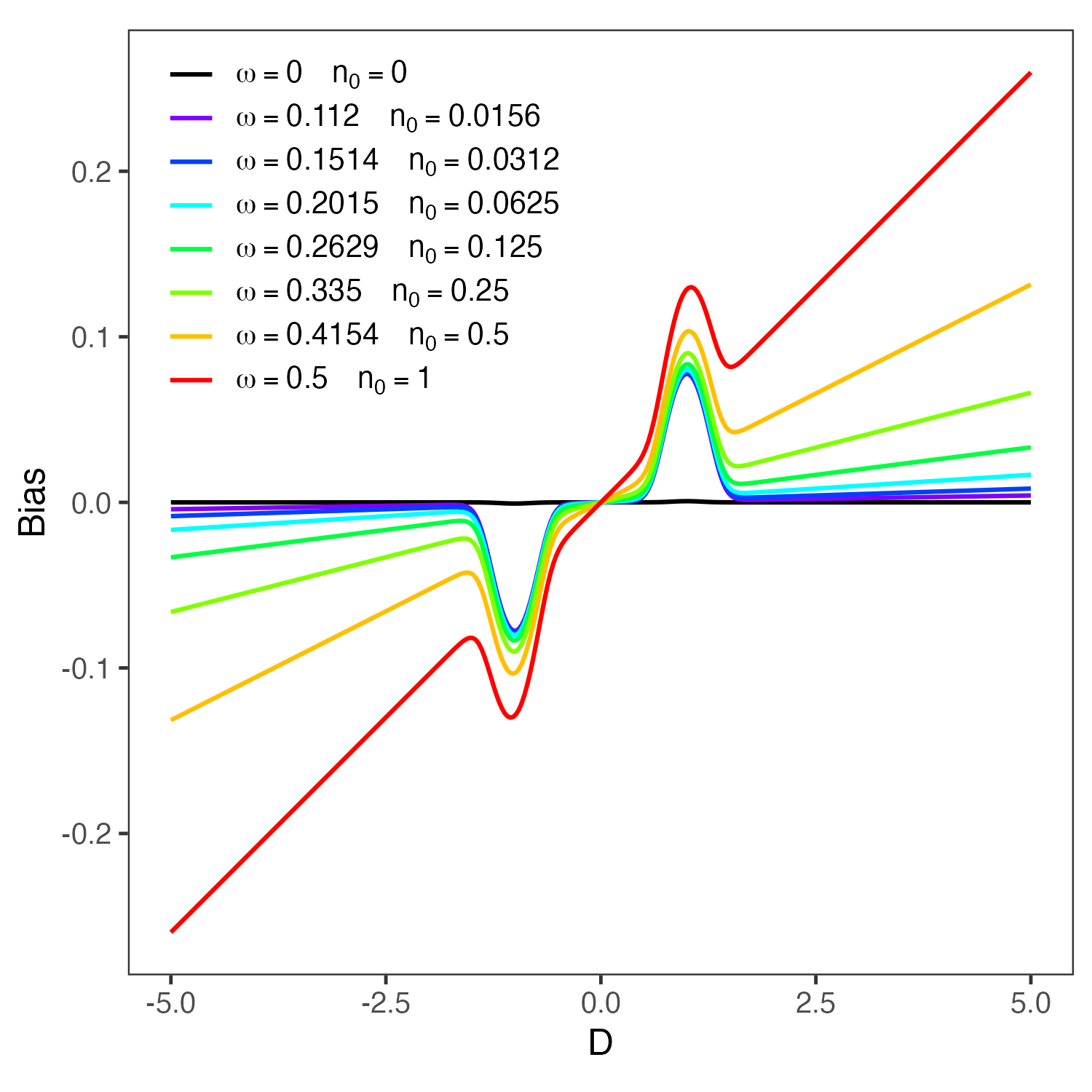}
        \label{bias.fig}}
	\end{subfigure}
    \hfill
	\begin{subfigure}[Variance]{
		\centering
		\includegraphics[width=0.3\textwidth]{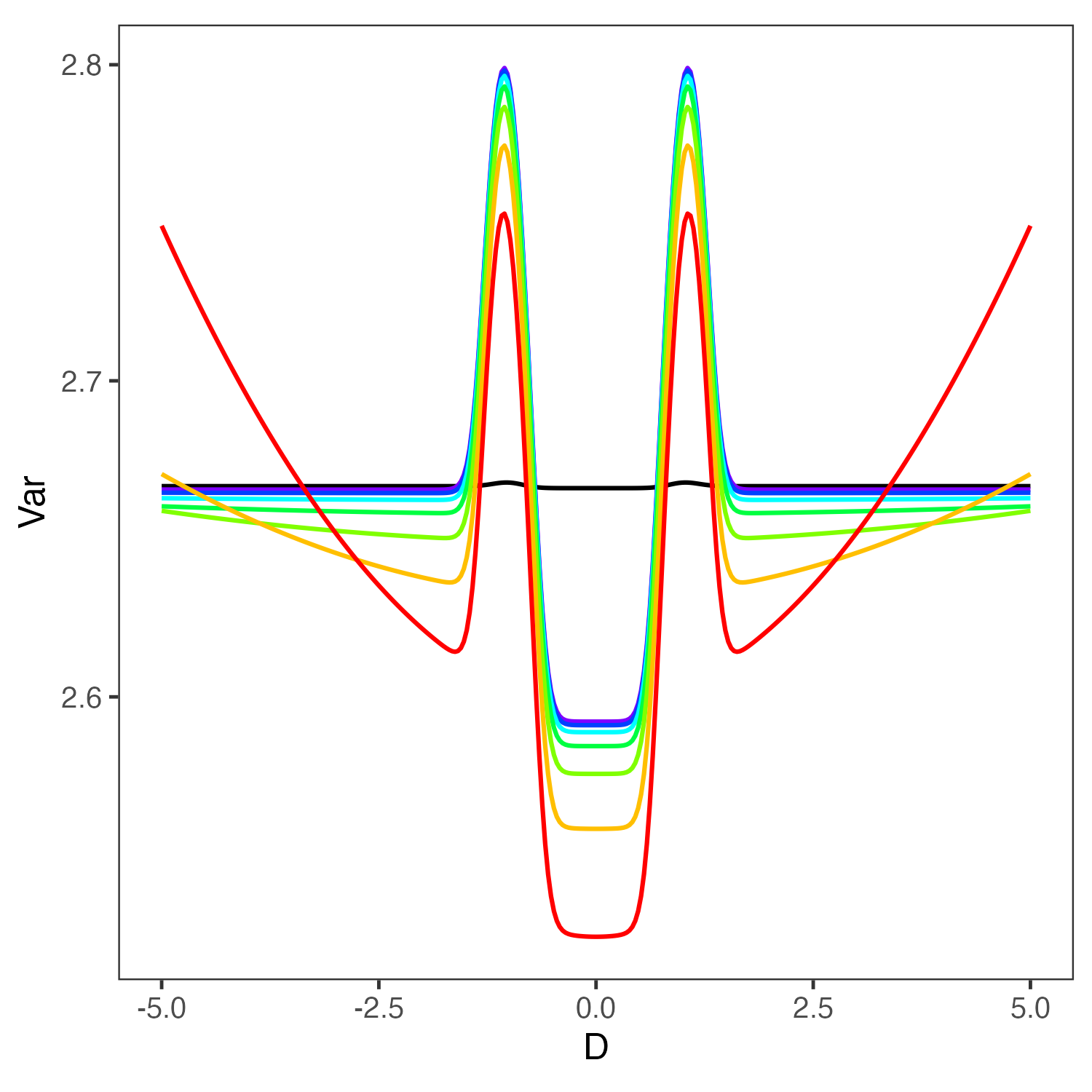}
        \label{variance.fig}}
	\end{subfigure}
    \hfill
	\begin{subfigure}[MSE]{
		\centering
		\includegraphics[width=0.3\textwidth]{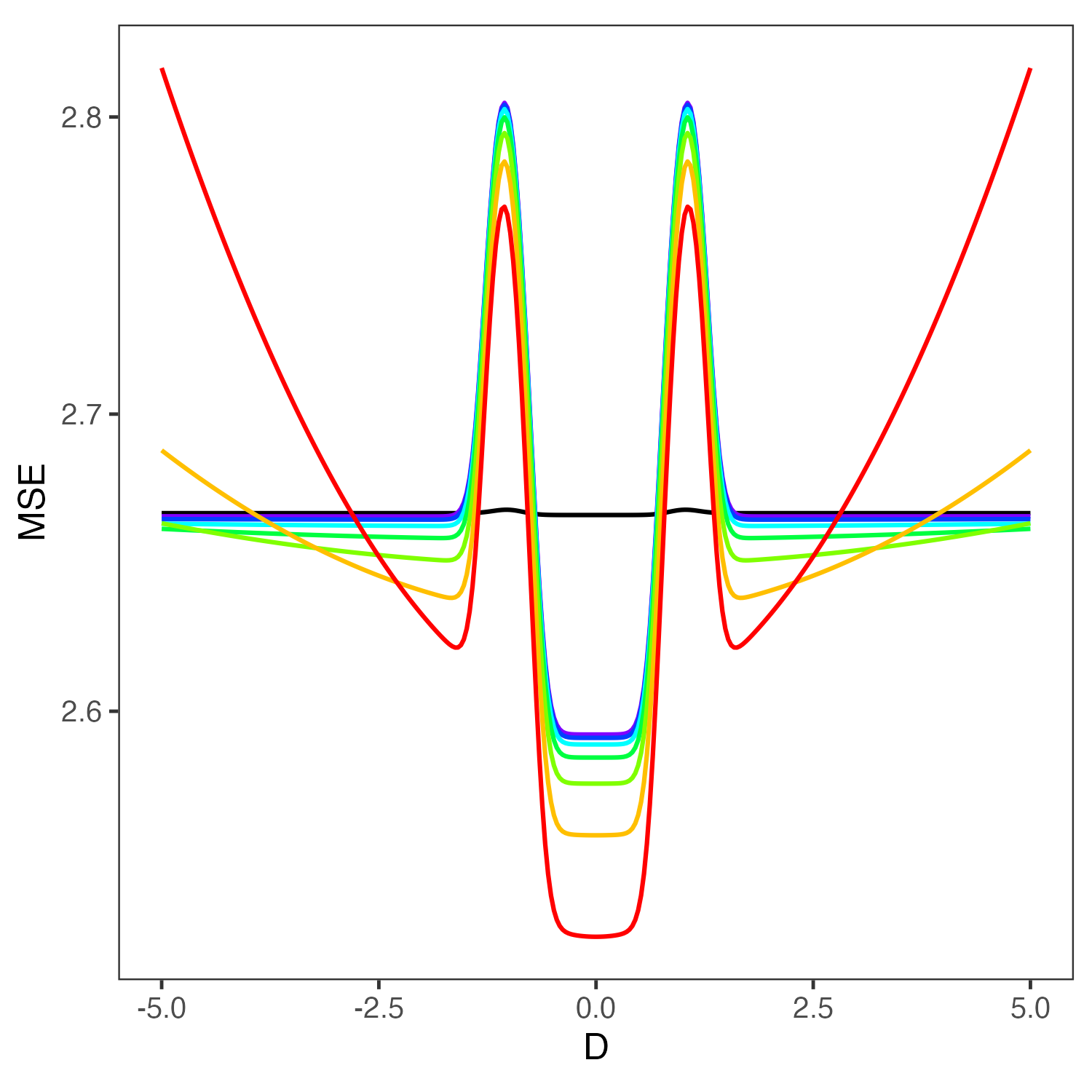}
        \label{MSE.fig}}
	\end{subfigure}
    
	\caption{Panel~(a): bias; Panel~(b): variance; Panel~(c): mean squared error, all computed using the posterior mean of the treatment effect parameter~$\delta$. Colors denote different pairs of~$(\omega, n_0)$, each corresponding to~$\beta^{*} = 5.83$.
}
    \label{diag.fig}
\end{figure}

\noindent
This application considers eight distinct RMPs, generated by varying the effective sample size of the control parameter, $n_0$, across the set $\{(\frac{1}{2})^k | k = 0, \dots, 7\}$, and the prior mixture weight, $\omega$, across the set $\{0, 0.5, 0.415, 0.335, 0.263, 0.201, 0.151, 0.112\}$. All considered pairs, excluding the first (representing an improper prior), belong to the level set $\beta(n_0, \omega) = 5.83$, thus exhibiting the shared posterior weight profile discussed in Section \ref{5.1}. For each RMP, type I error rate (Figure \ref{t1e.fig}) and power (Figure \ref{pow.fig}), are assessed, with power calculated for a treatment difference of $\delta^* = 0.31$. Posterior inference is evaluated using bias (Figure \ref{bias.fig}), variance (Figure \ref{variance.fig}), and mean squared error (MSE) (Figure \ref{MSE.fig}).  \vspace{0.2 cm} 

\noindent
For small to moderate prior-data conflicts, the power (Figure \ref{pow.fig}) and type I error rate (Figure \ref{t1e.fig}) curves overlap for all RMPs. This occurs because both variances and bias are comparable in these regions. Consequently, the posterior distributions of the treatment difference $\delta$ are similar across pairs, centered near $\delta = 0$ (for type I error rate) and $\delta = \delta^*$ (for power). This results in highly similar null hypothesis rejection rates for all RMPs.

%\begin{comment}
 
%\end{comment}

\noindent
Conversely, significant differences among the pairs emerge under large prior-data conflicts, where RMPs with weakly informative robustification components exhibit inflation (deflation) of both type I error rate and power for large positive (negative) drifts. However, this effect is attenuated for RMPs with less informative robustification components, practically disappearing when $n_0 < (\frac{1}{2})^6$. In these regions, substantial differences in bias among the RMPs impact type I error rate and power, which deviate considerably from their nominal levels for RMPs with more informative robustification components, while remaining near their nominal values for RMPs with less informative robustification components.

%\begin{comment}
    \begin{table}[H]
\centering
\begin{tabular}{cccccccccc}
\toprule
$\omega$ & $n_0$ & $\alpha_{max}$ & $\alpha(50)$ &  $\alpha^{\text{VAG}}_{avg}$ &  $\alpha^{\text{INF}}_{avg}$ &  $\alpha^{\text{RMP}}_{avg}$ & $\text{Pow}(0)$ & Sweet spot width \\
\midrule
0 & $10^{-100}$ & 0.0500 & 0.0500 & 0.0500 & 0.0500 & 0.0500  & 0.600 & 0.000 \\
0.500 & 1.000 & 0.168 & 0.9914 & 0.2955 & 0.0394 & 0.0492  & 0.803 & 0.207 \\
0.415 & 0.500 & 0.167 & 0.6478 & 0.1522 & 0.0397 & 0.0496  & 0.803 & 0.206 \\
0.335 & 0.250 & 0.166 & 0.2643 & 0.0785 & 0.0399 & 0.0498  & 0.802 & 0.207 \\
0.263 & 0.125 & 0.166 & 0.1278 & 0.0574 & 0.0399 & 0.0499  & 0.802 & 0.207 \\
0.201 & 0.062 & 0.166 & 0.0822 & 0.0520 & 0.0400 & 0.0499  & 0.802 & 0.207 \\
0.151 & 0.031 & 0.165 & 0.0645 & 0.0507 & 0.0400 & 0.0500  & 0.802 & 0.207 \\
0.112 & 0.016 & 0.165 & 0.0569 & 0.0503 & 0.0400 & 0.0500  & 0.802 & 0.207 \\
\bottomrule
\end{tabular}
\vspace{0.3cm}
\caption{Maximum type I error rate ($\alpha_{max}$), average type I error rate ($\alpha_{avg}$), power gain under no data-conflict $\text{Pow}(0)$ and width of the sweet spot for different couples of $(\omega, n_0)$, all corresponding to $\beta^{*}=5.83$.}
\label{table.OC}
\end{table}

%\end{comment}

\noindent
Table \ref{table.OC} summarizes key characteristics of the observed curves.  These include the maximum type I error rate inflation, $\alpha_{max}$, constrained to the interval $-5 < D < 5$ (a plausible response range); the power gain, $\text{Pow}(0)$, when the informative component of the RMP perfectly matches the control data; the type I error rate under extreme drift, $\alpha(50)$; the average type I error rate across different design priors (an improper prior, the informative component of the RMP, and the RMP itself); and the width of the ``sweet spot'' region \supercite{Viele2014}. The ``sweet spot'' is defined as the interval of $D$ values where type I error rate and Power are respectively below and above their nominal levels (5\% and 60\% in this application).

\vspace{0.2cm}\noindent
All considered $(\omega, n_0)$ pairs demonstrate comparable performance in terms of maximum type I error rate, $\alpha_{max}$, power gain, $\text{Pow}(0)$, and sweet spot width. However, a significant difference emerges when examining $\alpha(50)$.  This value is notably higher for RMPs with weakly informative robustification components (approaching 100\% for the UIP), progressively decreasing towards 5\% as the informativeness of the robustification component increases. \vspace{0.2 cm} \\  \noindent\noindent
\noindent
Averaging type I error rate across an improper prior distribution reveals a marked inflation for RMPs with weakly informative robustification components, as consequence of the asymptotic type I error rate increase discussed previously. The type I error rate decrease observed for negative drifts does not fully compensate for the inflation because the range of increase (from 5\% to 100\%) is considerably larger than the range of decrease (from 5\% to 0\%), leading to a greater weighting of the inflation in the averaging process. \vspace{0.2 cm} \\  \noindent\noindent
\noindent
Conversely, minimal differences are observed among pairs when averaging type I error rate across more informative priors, such as the informative component of the RMP or the RMP itself. These priors are concentrated around regions of small drifts, where all RMPs have practically identical type I error rate curves.  The type I error rate reduction exhibited by all RMPs in this region keeps the average type I error rate controlled at the nominal level (in the strong sense, when using the informative component or the RMP as the design prior). \vspace{0.2 cm} \\  \noindent\noindent
\noindent
In summary, RMPs with high-variance robustification components achieve comparable performance to those with weakly informative robustification components, while simultaneously mitigating type I error rate inflation.  This results in average type I error rate remaining below the nominal level when the RMP or its informative component are used as design priors (as demonstrated in Best et al. \supercite{Best2023}), but also controlled just slightly above the nominal level when improper priors are used; thus guaranteeing an higher overall protection to incorrect rejections of the the null hypothesis.

\subsection{Overcoming biases due to the specification of \textbf{$\mu_{\text{rob}}$}} \label{5.3}

\begin{figure}[H]
   \centering
   \includegraphics[width=0.9\linewidth]{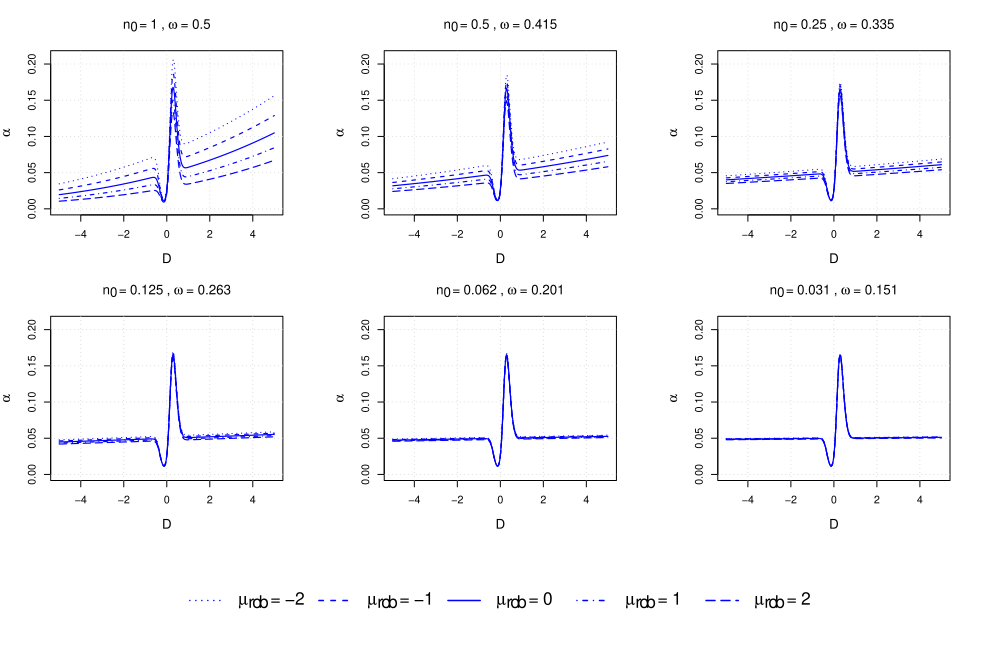}
   \caption{For each panel representing a different couples of $(\omega, n_0)$, type I error rate as a function of the prior-data conflict $D$ is displayed for five different values of the location of the robustification component $\mu_{\text{rob}}$.}
   \label{t1e.loc}
\end{figure}

Figure \ref{t1e.loc} investigate the influence of robustification component location on the type I error rate within the Robust Mixture Prior (RMP). For each of the first six $(\omega, n_0)$ pairs analyzed in Figure \ref{OC.fig} and Table \ref{table.OC}, five type I error rate and power curves (as functions of the drift parameter $D$) are presented, corresponding to variations in the robustification component location parameter, $\mu_{\text{rob}}$, across the set $\{-2, -1, 0, 1, 2\}$.

%\begin{comment}
%\end{comment}

\noindent
The figures demonstrate that for large $n_0$ values (e.g., UIP), operating characteristics exhibit high sensitivity to the location parameter $\mu_{\text{rob}}$. Consistently with what shown in Section \ref{3}, increasing $\mu_{\text{rob}}$ uniformly inflates both type I error rate curve, while decreasing $\mu_{\text{rob}}$ has the opposite effect. 
Conversely, as $n_0$ decreases (and accordingly $\sigma^2_{\text{rob}}$ increases), the impact of $\mu_{\text{rob}}$ on posterior inference diminishes, as evidenced by the substantial overlap of the type I error rate curves when $n_0 = 0.031$. The same behavior can be appreciated in the Power analysis in Figure \ref{pow.loc} of the supplementary material. 
%\vspace{0.2cm} \\
%This observation empirically validates Theorem \ref{theo.loc} and highlights the advantage of employing RMPs with high-variance robustification components.

\section{Hyper-parameters elicitation}
\label{6}

\subsection{On the interpretation of the prior weight}

The use of normal RMPs in practice necessitates the pre-specification of hyper-parameters: the robustification component location $\mu_{\text{rob}}$, the robustification component variance $\sigma^2_{\text{rob}}$, and the mixture weight $\omega$. Current practice often prioritizes default values for the two former parameters, centering the robustification component at the informative component mean ($\mu_{\text{rob}} = \mu_{\text{inf}}$) and selecting a unit-information robust variance \supercite{Schmidli2014}.  The mixture weight $\omega$ is then normally determined based on stakeholder or experts confidence in the data supporting the informative component. \\
This elicitation is typically driven by questions like \textit{``how much is the probability that historical data are relevant in the current setting?''} or \textit{``how much confidence do you have in historical data being representative of the current data?''}. For instance, high confidence (or high probability) might lead to $\omega = 0.9$, whereas low confidence might lead to $\omega = 0.3$. \vspace{0.2 cm} \\  \noindent\noindent
\noindent
While straightforward to communicate, this interpretation may disregard the crucial interplay between $\omega$ and $\sigma^2_{\text{rob}}$, significantly influencing RMP performance as it only concerns one parameter of the RMP, while it is argued above that they should be chosen in accordance with the variance of the robustification component. Furthermore, implies that the current choice of $\omega$ is unrelated to the choice of the robustification component. In fact, following the results above, we argue that the interpretation (and as a result the elicitation) of the weight should come together with the choice of the robustification component. \vspace{0.2 cm} \\  \noindent
We have proven in Section \ref{4.3} that the \textit{borrowing strength} $\beta^*$ is the key parameter influencing the borrowing profile of the RMP. This suggests that an equivalent prior degree of confidence in historical data should correspond to a lower $\omega$ for RMPs with a larger robustification component variance and a higher $\omega$ for RMPs with a smaller robustification component variance. As a consequence, we posit that $\omega$ should be viewed as a \textit{relative} confidence measure between the informative model $\pi_{\text{inf}}$ and the robust model $\pi_{\text{rob}}$, which specification should then depend on how informative the robustification component itself is. \vspace{0.2 cm} \\  \noindent
Given the suggested interpretation of $\omega$, we propose the following procedure for its elicitation.

\subsection{An approach for hyper-parameters elicitation}
A four-step elicitation approach is proposed: 
\begin{enumerate}
    \item Standard deviation of the robustification component of the RMP $\sigma_{\text{rob}}$ is set to a large value. A possible option is setting it to $\sigma_{\text{rob}}=1000\times s$, where $s$ represents the standard deviation of the considered endpoint (note that even higher values can be used, but as demonstrated above they will have no impact on the inference).
    \item The location of the robustification component $\mu_{\text{rob}}$ is set equal to the location of the informative component $\mu_{\text{inf}}$.
    \item Clinicians are asked to determine an ``equipoise drift'' value $d^*$, representing the potentially observed control response that would induce maximum uncertainty regarding the relevance of historical data. Prompting questions could be: \textit{``At what control response value would you be 50\% confident that the historical component is relevant for the current trial and 50\% that it is not?''} or \textit{``At what control response value would you suspect a systematic difference between historical and concurrent control data?''}.
    \item Once specified $\sigma_{\text{rob}}$ and $d^*$, the prior odds $\Omega$ is obtained such that $\Tilde{\Omega}(d^{*}+\mu_{\text{inf}})=1$ (or equivalently $\tilde{\omega}=0.5$), inverting equation (\ref{update normal}) as follows:
    \begin{equation}
        \Omega = \frac{R}{\exp\!\left\{
- \frac{d^{*2}}{2v_{\text{inf}}^2} + 
\frac{(x_c-\mu_{\text{rob}})^2}{2R^2 v_{\text{inf}}^2}
\right\}}
    \end{equation}
    and accordingly the prior weight is retrieved as $\omega = \frac{\Omega}{1+\Omega}$.
    
\end{enumerate}

\noindent
Our hyper-parameter selection routine combines the benefits of RMPs with large variance robustification components and expert interaction. Moreover, while elicitation of the mixture weight $\omega$ poses challenges due to its complex interpretability, elicitation on the drift scale offers straightforward interpretation, thus justifying the approach.

\section{Beta-Binomial case}
\label{7}

\subsection{Beta Robust Mixture Prior}
Let us now consider the setting in which a RCT is performed with a binary outcome so that the total number of responses is $X_c \sim \text{Bin} \left( \theta_c, n_c \right)$, where $n_c$ is the number of patients allocated to the control arm and $\theta_c \in (0,1)$ represents the response parameter on the probability scale. 
\vspace{0.2cm} \\
\noindent
The Robust Mixture Prior in this case can be chosen as a mixture of two Beta distribution, namely $\text{Beta} \left(a_{\text{inf}}, b_{\text{inf}} \right)$ for the informative component and $\text{Beta} \left(a_{\text{rob}}, b_{\text{rob}} \right)$ for the robustification component. Then the prior predictive density of the data is a Beta-Binomial, namely 

\begin{equation}
\label{predictive_beta}
    f\left(x_c | \pi_{\star} \right) = \binom{n_c}{x_c} \frac{B \left( a_{\star} + x_c, b_{\star} + n_c -x_c \right)}{B \left( a_{\star}, b_{\star} \right)} \; \; \; \; \; \; \; \; \; \; \; \; \; \; \; \; \star= \{ \text{inf, rob} \} 
\end{equation}

\noindent
where $x_c \in \left(0,n_c \right)$ is the observed number of responders in the control arm and $B(\cdot)$ represents the Beta function. Working out with the Gamma function expression of the Beta function, it follows that the odds update of Equation \eqref{Odds update} can be expressed in this case as

\begin{equation} \label{odds_beta}
    \Omega \left( x_c \right) = \beta\left(\omega, a_{\text{rob}}, b_{\text{rob}} \right) \times \frac{B \left( a_{\text{inf}} + x_c, b_{\text{inf}} + n_c - x_c\right)}{B \left( a_{\text{rob}} + x_c, b_{\text{rob}} + n_c - x_c\right) B \left(a_{\text{inf}}, b_{\text{inf}} \right)},
\end{equation}

\noindent
where the function $\beta\left(\omega, a_{\text{rob}}, b_{\text{rob}} \right)$ can be expressed as 

\begin{equation}\label{beta_fun_betabinom}
    \beta\left(\omega, a_{\text{rob}}, b_{\text{rob}} \right) = \Omega \cdot B \left(a_{\text{rob}}, b_{\text{rob}} \right)
\end{equation}

\noindent
Note that although $a_{\text{rob}}$ and $b_{\text{rob}}$ may differ, setting them equal and small is a reasonable choice when aiming to represent limited prior knowledge. In common practice, specifications such as $\mathrm{Beta}(1,1)$ or $\mathrm{Beta}(0.5,0.5)$ (Jeffreys prior) are typically employed for this purpose.

\subsection{The Lindley's paradox in the Beta-Binomial case}

Similarly to the normal case, also in the Beta-Binomial case the phenomenon of the Lindley's paradox occurs when a large variance distributions is used as a robust component of the RMP. Specifically, this happens - for a fixed $\omega$ - when the parameter of the Beta distribution related to the robust component approaches 0, because $\Gamma \left( 0^+\right) \rightarrow +\infty$ and accordingly following Equation \eqref{beta_fun_betabinom} the posterior odds goes to $+\infty$ and accordingly the posterior weights $\omega$ goes to 1. Similarly to what done in the normal case in Theorem \ref{theo.lindley}, in Theorem \ref{theo.lindley.binom} we show that this behavior is due to the hidden underlying assumption that the mixture weight $\omega$ is fixed and independent on the choice of $a_{\text{rob}}$ and $b_{\text{rob}}$. We find that relaxing this assumption, effectively prevents from the occurring of Lindley's paradox. 

\begin{theorem} \label{theo.lindley.binom}
    Consider a binomial random variable $X_c \sim \text{Bin} \left(\theta_c, n_c \right)$, and assume a RMP is used for the parameter $\theta_c$, namely $\pi_c(\theta_c) = \omega \pi_\text{inf}(\theta_c)+(1-\omega) \pi_\text{rob}(\theta_c)$, where $\pi_\text{inf}(\theta_c)$ and $\pi_\text{rob}(\theta_c)$ are the PDF of Beta distributed random variables with parameters $a_{\text{inf}}$, $b_{\text{inf}}$ and $a_{\text{rob}} = b_{\text{rob}} = \varepsilon$, respectively. The following hold:
    \begin{enumerate}
        \item if $\Omega < + \infty$, then
        \begin{equation*}
        \lim_{\varepsilon \rightarrow 0} \Tilde{\omega} \left(x_c, \pi_\text{inf}(\theta_c), \pi_\text{rob}(\theta_c), \omega \right) = 1 \; \; \; \; \; \; \; \forall x_c \in \left( 0, n_c \right)
        \end{equation*}
        
        \item if $\Omega \sim O \left( \varepsilon \right)$ for $ \varepsilon \rightarrow 0$, then
        \begin{equation*}
            \lim_{\varepsilon \rightarrow 0} \Tilde{\omega} \left(x_c, \pi_\text{inf}(\theta_c), \pi_\text{rob}(\theta_c), \omega \right) \neq 1 \; \; \; \; \; \; \; \forall x_c \in \left( 0, n_c \right)
        \end{equation*}    
    \end{enumerate}
\end{theorem}

\noindent
A formal proof of Theorem 4 can be found in the Supplementary material. \\
\noindent
The preceding theorem demonstrates that Lindley's paradox arises, as the parameters of the robust component of the RMP approaches zero, when the prior weight $\omega$ (or prior odds $\Omega$) is fixed independently of the parameters of the robust component. Conversely, if $\omega$ and $a_{\text{rob}} = b_{\text{rob}} = \varepsilon$ are jointly selected such that the prior odds $\Omega$ remain of the same order of magnitude as the parameters of the robust component, namely  $\Omega \sim O(\varepsilon)$, then Lindley's paradox is avoided.  
\vspace{0.2cm} \\
\noindent
This occurs because, as $\varepsilon \to 0$, the posterior odds $\tilde{\Omega}$ can be expressed following Equations \eqref{odds_beta} and \eqref{beta_fun_betabinom} as
\begin{equation}
\tilde{\Omega}(x_c; \omega, \varepsilon) = \beta(\omega, \varepsilon) \times \frac{B \left( a_{\text{inf}} + x_c, b_{\text{inf}} + n_c - x_c\right)}{B \left( x_c, n_c - x_c\right) B \left(a_{\text{inf}}, b_{\text{inf}} \right)} \; ,
\end{equation}
\noindent
where the influence of the RMP on the posterior odds is entirely captured by the function $\beta(\omega, \varepsilon)$ defined in Equation \eqref{beta_fun_betabinom}. It follows that, similarly to what shown in the normal case, all combinations of $\omega$ and $\varepsilon$ yielding the same $\beta(\omega, \varepsilon) = \beta^*$ share the same “borrowing profile”, resulting in identical posterior odds and posterior weights $\tilde{\omega}$ for any observed number of responders $x_c$.  
\vspace{0.2cm} \\
\noindent
The parameter $\beta^*$ governs the RMP's flexibility in borrowing information across the $x_c$ space, determining the rate at which posterior weights decrease in the presence of prior-data conflict.  
\vspace{0.2cm} \\
\noindent
It is important to note that, while these pairs $(\omega, \varepsilon)$ yield identical posterior weights, posterior inference for $\theta_c$ could in principle differ across RMPs due to variations in the robust posterior component $g_\text{rob}(\theta_c | x_c, \pi_\text{rob})$ arising from different choices of $\varepsilon$. However, as $\varepsilon \to 0$, the posterior distribution related to the robust component of the RMP tends to lose its dependence on the prior parameters, thus leading to similar inference for $\theta_c$ across all such pairs.

\subsection{Practical Considerations}

In the Supplementary Material, the results presented in Section 7 are validated through a numerical investigation. Specifically, we considered a randomized controlled trial (RCT) in which $n_c = 100$ patients are assigned to the control arm, while $n_t = 200$ patients are allocated to the treatment arm. The number of responses in each arm follows a binomial distribution $X_{*} \sim \text{Bin}(\theta_{*}, n_{*}), \; * = \{c,t\}.$
\vspace{0.2cm} \\
\noindent
A Jeffreys prior, $\text{Beta}(0.5,0.5)$, is used for the treatment parameter $\theta_t$, whereas various robust mixture priors (RMPs) are explored as prior distributions for the control parameter $\theta_c$. The informative component of the RMP is fixed to $\text{Beta}(50,50)$, reflecting a prior knowledge on the control parameter being close to $\theta_c=0.5$. The success rule is the same expressed in Equation \eqref{success RCT}, where $\delta$ represents the log odds ratio corresponding to the two parameters, namely $\delta=\log\left( \frac{\theta_t (1-\theta_c)}{\theta_c (1-\theta_t)}\right)$.
\vspace{0.2cm} \\
\noindent
Analogously to the normal case, Figure~\ref{3d_beta} illustrates how the posterior weights vary as a function of the observed number of responses in the control arm, when the prior weight $\omega$ and the parameters of the robust component of the RMP, $a_{\text{rob}} = b_{\text{rob}}$, are jointly chosen to satisfy the condition $\beta^* = 12.56$. Notice that this value has been arbitrarily selected so to include the pair $\omega=0.8$, $a_{\text{rob}}=b_{\text{rob}}=0.5$, so that $\beta^* = \beta(0.8, 0.5)= \frac{0.8}{1-0.8} \cdot B(0.5,0.5)$.

\begin{figure}[H]
    \centering

    % Immagine sinistra
    \includegraphics[width=0.45\textwidth]{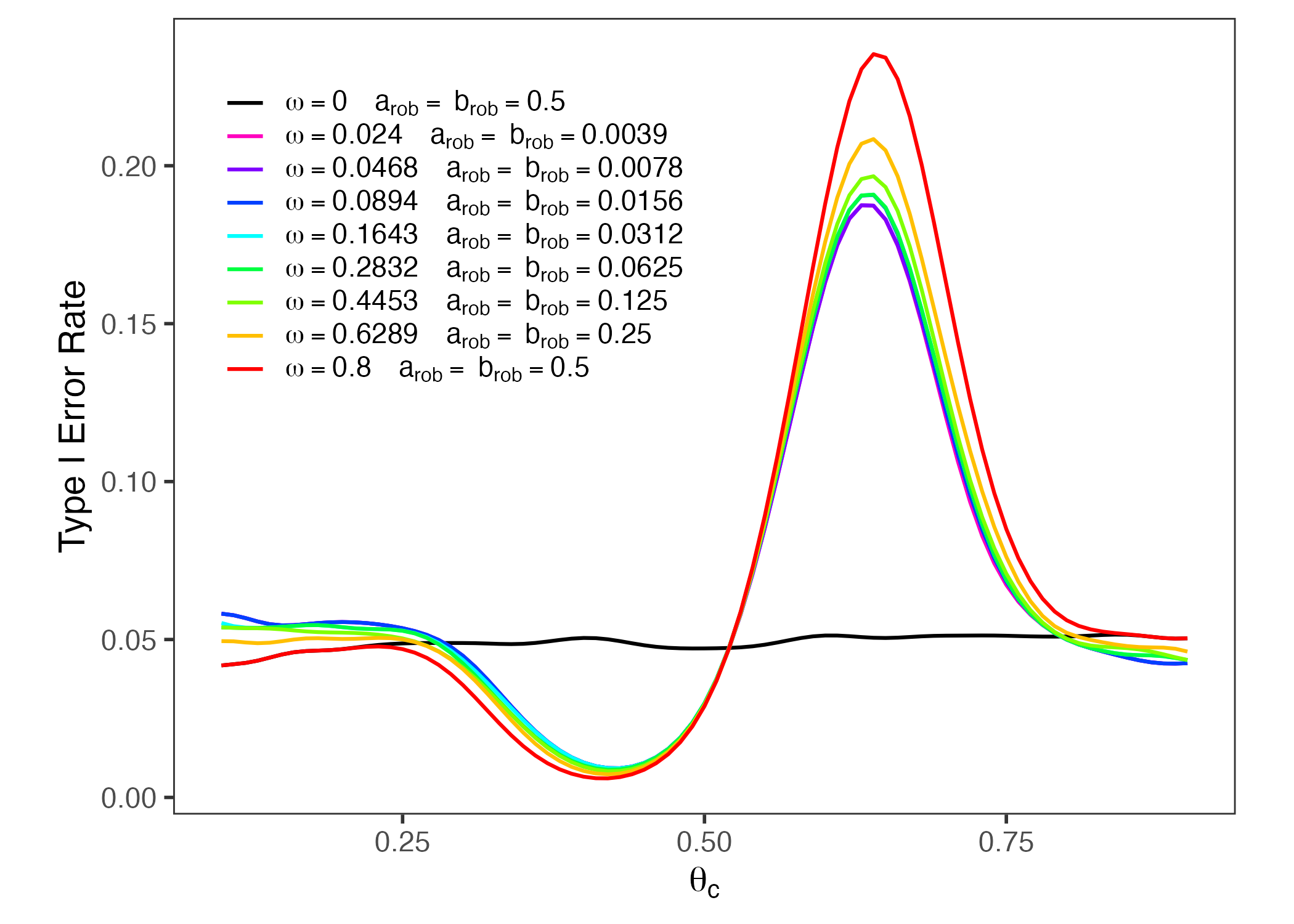}
    \hfill
    % Immagine destra
    \includegraphics[width=0.45\textwidth]{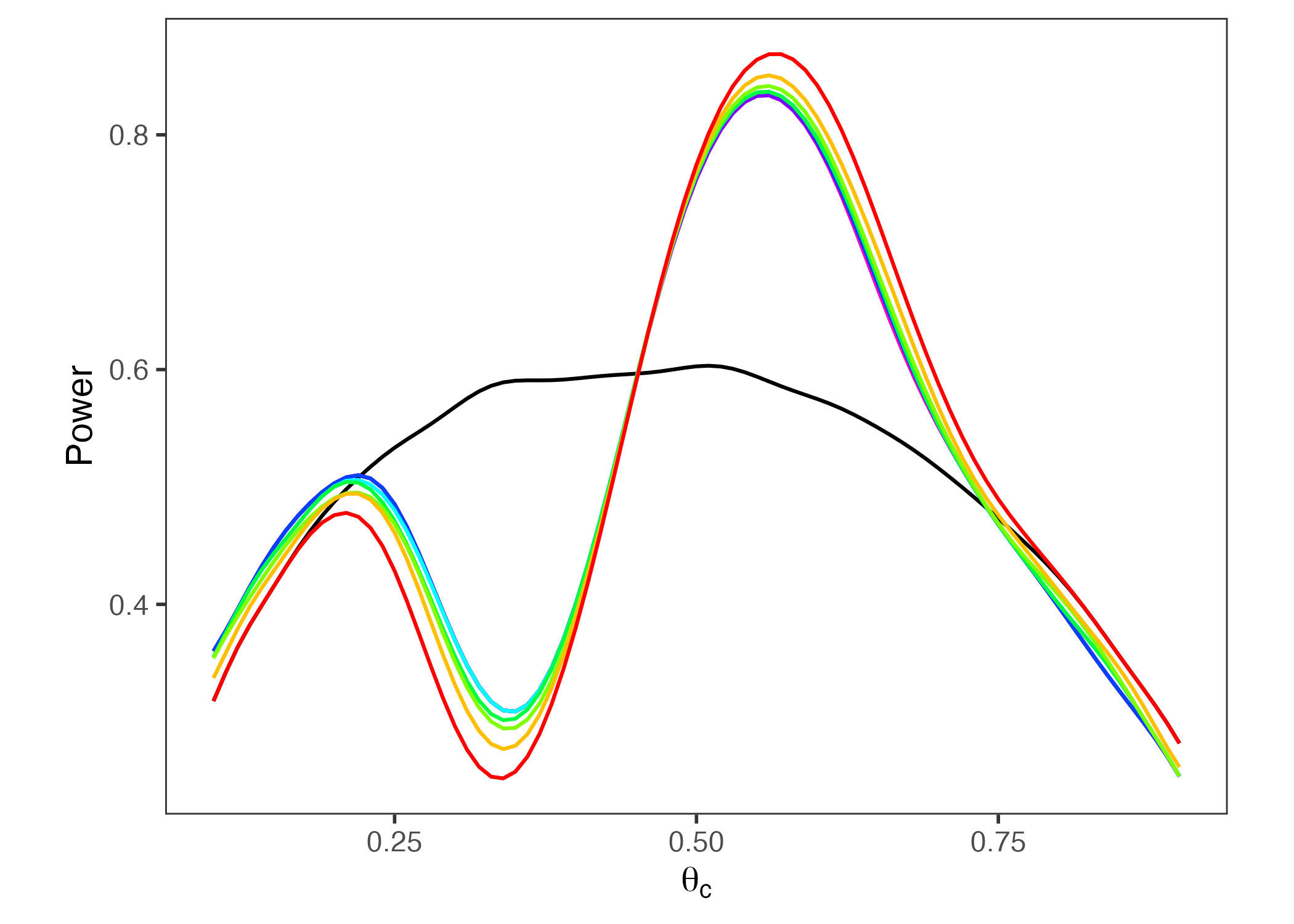}

    \caption{Panel~(a): Type~I error rate}; Panel~(b): power under a target log-odds ratio~$\delta^{*} = 0.47$, both evaluated in the Beta--Binomial setting. Colors indicate different pairs of~$(\omega, a_{\text{rob}}=b_{\text{rob}})$ corresponding to~$\beta^* = 12.56$.

    \label{OC.fig.beta}
\end{figure}

\vspace{0.2cm}
\noindent
The figure shows that, for all parameter pairs satisfying $\beta^* = 12.56$, the variation of the posterior weights $\tilde{\omega}$ with respect to the number of control responses $x_c$ is closely aligned. This indicates that all such RMPs exhibit the same borrowing profile, and particularly that borrowing is possible even when $a_{\text{rob}}$ and $b_{\text{rob}}$ are very small, thus confirming that the Lindley's paradox can be effectively avoided provided a joint selection of the pair $(\omega, a_{\text{rob}}= b_{\text{rob}})$.
\vspace{0.2cm} \\
\noindent
This behavior is further confirmed by examining the type I error rate and power plots in Figure~\ref{OC.fig.beta}, as well as the bias, variance, and mean squared error plots in Figure~\ref{S6.fig.beta}.

\vspace*{\fill} % inizio centratura verticale
\vspace*{\fill} % fine centratura verticale

\noindent
In these figures, eight pairs $(\omega, a_{\text{rob}} = b_{\text{rob}})$ satisfying $\beta^* = 12.56$ are shown, and the operating characteristics corresponding to different RMPs are displayed across the true control parameter $\theta_c \in (0.1,0.9)$. In particular, the curves corresponding to different pairs $(\omega, a_{\text{rob}} = b_{\text{rob}})$ follow very similar trends across the $\theta_c$ range. A near-complete overlap is observed for pairs with $a_{\text{rob}} = b_{\text{rob}} < 0.1$ across the parameter space, while some deviations occur in regions of moderate prior-data conflict, i.e., when more informative Beta priors are employed as the robust component of the RMP. For instance, using a $\text{Beta}(0.5,0.5)$ prior produces similar OCs in regions of minor drift, but the maximum type I error increases noticeably (approximately 6\% higher) relative to RMPs with weaker robust components, due to higher bias in regions of intermediate conflict.
\vspace{0.2cm} \\
\noindent
Consistent with the normal case, we conclude that employing quasi non-informative Beta distributions as the robust component in the Beta RMP is feasible without inducing Lindley’s paradox, provided that the prior weight $\omega$ and the parameters of the robust component are jointly selected. Moreover, using weakly informative robust components mitigates bias in regions of the parameter space where type I error inflation is most pronounced, thus offering greater protection against potential inflation arising from moderate drift between concurrent and historical data.

\vspace{0.2cm}
\noindent
Finally, it is noteworthy that, in the Beta-Binomial setting, asymptotic type I error inflation is not a concern, as the extent of prior-data conflict is inherently bounded by the domain of the parameter $\theta_c$.

\section{Extension to a Mixture Informative component}\label{8}

The framework introduced in this paper can be further extended to the case in which the informative component of the Robust Mixture Prior (RMP) is itself modeled as a mixture of distributions, such as Beta or Normal, depending on the context. 
\vspace{0.2cm} \\
\noindent
Let the informative component of the RMP be expressed as
\begin{equation}
    \pi_{\mathrm{inf}}(\theta_c) = \sum_{k=1}^{K} \xi_k \, \pi_{\mathrm{inf}}^{(k)},
\end{equation}
\noindent
where $\sum_{k=1}^{K} \xi_k = 1$. Denote by $\omega$ the weight assigned to the informative component and by $1 - \omega$ the weight assigned to the robust component. The overall RMP can then be represented as a mixture of $K+1$ components:
\begin{equation}
    \pi_{c}(\theta_c) = \sum_{k=1}^{K} \omega \, \xi_k \, \pi_{\mathrm{inf}}^{(k)} 
    + (1 - \omega)\, \pi_{\mathrm{rob}}.
\end{equation}

\noindent
Define $\eta_k = \omega \, \xi_k$ for $k = 1, \dots, K$ and $\eta_{K+1} = 1 - \omega$. 
Let $\Omega_k = \eta_k / (1 - \eta_k)$ denote the odds associated with the $k$-th component of the RMP. An extension of Equation~\ref{Odds update} to this setting, expressed in terms of the reciprocal of the odds rather than the odds themselves (for convenience), can be written as

\begin{equation} \label{odds_mix}
\tilde{\Omega}^{-1}_h(x_c)
= 
\displaystyle
\sum_{\substack{k = 1 \\ k \neq h}}^{K}
\frac{%
  \xi_k\, f\!\left(x_c \mid \pi_{\mathrm{inf}}^{(k)}\right)
}{%
  \xi_h\, f\!\left(x_c \mid \pi_{\mathrm{inf}}^{(h)}\right)
}
\;+\;
\frac{1}{\xi_h}\,
\Omega_{K+1}^{-1}\,
\frac{%
  f\!\left(x_c \mid \pi_{\mathrm{rob}}\right)
}{%
  f\!\left(x_c \mid \pi_{\mathrm{inf}}^{(h)}\right)
} \quad \quad h=1,\dots,K
\end{equation}

\noindent
and the posterior weight related to the robust component can be retrieved as $\Tilde{\eta}_{K+1} = 1-\sum_{k=1}^{K} \Tilde{\eta}_k$. Note that Equation \eqref{odds_mix} reduces to Equation \eqref{Odds update} when $K=1$.
\vspace{0.2cm} \\
\noindent
It is worth noting that the first summation term in the above expression does not depend on the prior weights assigned to the informative and non-informative components, but only on the fixed weights $\xi_k$ associated with each element of the informative part of the RMP. Moreover, it is independent of the specification of the robust component of the RMP. The reciprocal of the second term, in contrast, coincides with Equation~\ref{Odds update}, rescaled by a component-specific factor $\xi_h$. Consequently, the asymptotic decomposition derived in the previous sections (for both the continuous and binary cases) remains valid, and the proposed methodology can be seamlessly extended to the mixture-based framework.

\section{Discussion}
\label{9}

Robust Mixture Priors (RMPs) are a prominent dynamic borrowing approach used to incorporate historical control data in the analysis of a current randomized trial. However, specifying parameters for the RMP components, particularly the robustification component and mixture weights, presents a challenge, as these parameters strongly influence posterior inferences. While improper normal distributions may seem intuitive for the robustification component, their use has been discouraged due to the potential for Lindley's paradox, prompting a preference for weakly informative priors. Employing the unit-information prior (UIP) \supercite{Schmidli2014} has become common; nevertheless, this choice remains somewhat arbitrary and context-dependent \supercite{Callegaro2023}. Specifically, concerns have been raised regarding the UIP's potential over-informativeness in trials with limited sample sizes \supercite{Weru2024}, as well as the theoretical unbounded type I error rate in unbalanced trials using UIP \supercite{Best2023}. 

\vspace{0.2cm} 
\noindent
In this article, we demonstrate, for both normal and binary endpoints, that jointly eliciting the mixture weight and the hyperparameters of the robustification component within a Robust Mixture Prior (RMP) framework effectively mitigates Lindley's paradox, even when using arbitrarily large variances. 

\vspace{0.2cm} 
\noindent
This approach offers several practical advantages. In the normal case, it practically eliminates the impact of the location of the robustification component and prevents asymptotic type I error rate inflation in unbalanced trials, which is a critical regulatory consideration. While asymptotic inflation does not occur in balanced trials, these scenarios are of limited practical interest, as the main goal of borrowing is to reduce sample size on the control arm.

\vspace{0.2cm} 
\noindent
For binary endpoints, asymptotic type I error inflation does not occur due to the natural bounds of the probability parameter (0 to 1). Nevertheless, employing a large-variance robustification component (i.e., a Beta distribution with parameters approaching 0) has been shown to reduce the maximum type I error inflation compared to the commonly used Jeffreys prior.

\vspace{0.2cm} 
\noindent
We illustrate these properties through a proof-of-concept case study. Additionally, we propose a novel routine for selecting hyperparameters that combines a large-variance robustification component with an expert opinion-driven prior weight, $\omega$.

\vspace{0.2cm} 
\noindent
We further extend the methodology to the setting where the informative component of the RMP itself is a mixture of normal distributions, enhancing the flexibility of the approach.

\vspace{0.2cm} 
\noindent
Importantly, the insights derived from this work are general and extend to any framework employing a Robust Mixture Prior (RMP). The demonstrated interplay between the prior weight $\omega$ and the robustification component $\pi_{\text{rob}}$ is not limited to the specific implementation proposed here but is also relevant to other approaches that rely on RMPs, including those based on empirical Bayes formulations such as the EB-rMAP \supercite{Zhang2023} and the SAM prior \supercite{Yang2023}. Consequently, our findings provide a unifying perspective that can inform the specification and calibration of RMP-based borrowing mechanisms across diverse methodological frameworks.

\vspace{0.2cm} 
\noindent
Although the mathematical results could, in principle, be extended to one-arm trials where borrowing is performed on the treatment effect scale, exploring this application is beyond the scope of the current study. We leave the investigation of one-arm trial extensions and the evaluation of whether similar advantages hold in practice as future work.

%\begin{table}
% \caption{Sample table title}
%  \centering
%  \begin{tabular}{lll}
%   \toprule
%   \multicolumn{2}{c}{Part}        %           \\
%    \cmidrule(r){1-2}
%    Name     & Description     & Size ($\mu$m) \\
%    \midrule
%    Dendrite & Input terminal  & $\sim$100     \\
%    Axon     & Output terminal & $\sim$10      \\
 %   Soma     & Cell body       & up to $10^6$  \\
%  \bottomrule
% \end{tabular}
%  \label{tab:table}
%\end{table}
\section*{Acknowledgments}
This work was supported by Institut de Recherches Internationales Servier. The results reported herein are part of a collaboration between Servier, Saryga, and P. Mozgunov  whose research is supported by the National Institute for Health and Care Research (NIHR Advanced Fellowship, Dr Pavel Mozgunov, NIHR300576). The views expressed in this publication are those of the authors and not necessarily those of the NHS, the National Institute for Health and Care Research or the Department of Health and Social Care (DHCS). P Mozgunov received funding from UK Medical Research Council (MC UU 00040/03). M Gasparini received funding from MUR – M4C2 1.5 of PNRR funded by the European Union - NextGenerationEU (Grant agreement no. ECS00000036).

%\bibliographystyle{unsrt}  
%\bibliography{references}

\printbibliography

@article{Weru2024,
   author = {Vivienn Weru and Annette Kopp-Schneider and Manuel Wiesenfarth and Sebastian Weber and Silvia Calderazzo},
   month = {12},
   title = {Information borrowing in Bayesian clinical trials: choice of tuning parameters for the robust mixture prior},
   year = {2024},
}

@article{Hobbs2011,
   author = {Brian P Hobbs and Bradley P Carlin and Sumithra J Mandrekar and Daniel J Sargent},
   doi = {10.1111/j.1541-0420.2011.01564.x},
   issn = {1541-0420},
   issue = {3},
   journal = {Biometrics},
   month = {9},
   pages = {1047-56},
   pmid = {21361892},
   title = {Hierarchical commensurate and power prior models for adaptive incorporation of historical information in clinical trials.},
   volume = {67},
   year = {2011},
}

@article{Rover,
   author = {Christian Röver and Simon Wandel and Tim Friede},
   doi = {10.1002/sim.7991},
   issn = {0277-6715},
   issue = {4},
   journal = {Statistics in Medicine},
   month = {2},
   pages = {674-694},
   title = {Model averaging for robust extrapolation in evidence synthesis},
   volume = {38},
   year = {2019},
}

@article{Schmidli2014,
   author = {Heinz Schmidli and Sandro Gsteiger and Satrajit Roychoudhury and Anthony O'Hagan and David Spiegelhalter and Beat Neuenschwander},
   doi = {10.1111/biom.12242},
   issn = {0006-341X},
   issue = {4},
   journal = {Biometrics},
   month = {12},
   pages = {1023-1032},
   title = {Robust meta‐analytic‐predictive priors in clinical trials with historical control information},
   volume = {70},
   year = {2014},
}

@article{VanRosmalen,
   author = {Joost van Rosmalen and David Dejardin and Yvette van Norden and Bob Löwenberg and Emmanuel Lesaffre},
   doi = {10.1177/0962280217694506},
   issn = {0962-2802},
   issue = {10},
   journal = {Statistical Methods in Medical Research},
   month = {10},
   pages = {3167-3182},
   title = {Including historical data in the analysis of clinical trials: Is it worth the effort?},
   volume = {27},
   year = {2018},
}

@article{Viele2014,
   author = {Kert Viele and Scott Berry and Beat Neuenschwander and Billy Amzal and Fang Chen and Nathan Enas and Brian Hobbs and Joseph G. Ibrahim and Nelson Kinnersley and Stacy Lindborg and Sandrine Micallef and Satrajit Roychoudhury and Laura Thompson},
   doi = {10.1002/pst.1589},
   issn = {1539-1604},
   issue = {1},
   journal = {Pharmaceutical Statistics},
   month = {1},
   pages = {41-54},
   title = {Use of historical control data for assessing treatment effects in clinical trials},
   volume = {13},
   year = {2014},
}

@article{Callegaro2023,
   author = {Andrea Callegaro and Nicholas Galwey and Juan J Abellan},
   doi = {10.1093/biostatistics/kxab040},
   issn = {1465-4644},
   issue = {2},
   journal = {Biostatistics},
   month = {4},
   pages = {443-448},
   title = {Historical controls in clinical trials: a note on linking Pocock’s model with the robust mixture priors},
   volume = {24},
   year = {2023},
}

@article{Ibrahim2015,
   author = {Joseph G Ibrahim and Ming-Hui Chen and Yeongjin Gwon and Fang Chen},
   doi = {10.1002/sim.6728},
   issn = {1097-0258},
   issue = {28},
   journal = {Statistics in medicine},
   month = {12},
   pages = {3724-49},
   pmid = {26346180},
   title = {The power prior: theory and applications.},
   volume = {34},
   year = {2015},
}

@article{Mutsvari2016,
   author = {Timothy Mutsvari and Dominique Tytgat and Rosalind Walley},
   doi = {10.1002/pst.1722},
   issn = {1539-1604},
   issue = {1},
   journal = {Pharmaceutical Statistics},
   month = {1},
   pages = {28-36},
   title = {Addressing potential prior‐data conflict when using informative priors in proof‐of‐concept studies},
   volume = {15},
   year = {2016},
}

@article{Fougeray2024,
   author = {Ronan Fougeray and Loïck Vidot and Marco Ratta and Zhaoyang Teng and Donia Skanji and Gaëlle Saint‐Hilary},
   doi = {10.1002/pst.2410},
   issn = {1539-1604},
   journal = {Pharmaceutical Statistics},
   month = {7},
   title = {Futility Interim Analysis Based on Probability of Success Using a Surrogate Endpoint},
   year = {2024},
}

@article{Best2023,
   author = {Nicky Best and Maxine Ajimi and Beat Neuenschwander and Gaëlle Saint-Hilary and Simon Wandel},
   doi = {10.1080/19466315.2024.2342817},
   issn = {1946-6315},
   issue = {2},
   journal = {Statistics in Biopharmaceutical Research},
   month = {4},
   pages = {183-196},
   title = {Beyond the Classical Type I Error: Bayesian Metrics for Bayesian Designs Using Informative Priors},
   volume = {17},
   year = {2025}
}

@article{Yang2023,
   author = {Peng Yang and Yuansong Zhao and Lei Nie and Jonathon Vallejo and Ying Yuan},
   doi = {10.1111/biom.13927},
   issn = {15410420},
   journal = {Biometrics},
   publisher = {John Wiley and Sons Inc},
   title = {SAM: Self-adapting mixture prior to dynamically borrow information from historical data in clinical trials},
   year = {2023},
}

@article{Zhang2023,
   author = {Hongtao Zhang and Yueqi Shen and Judy Li and Han Ye and Alan Y. Chiang},
   doi = {10.1002/pst.2315},
   issn = {15391612},
   issue = {5},
   journal = {Pharmaceutical Statistics},
   month = {9},
   pages = {846-860},
   pmid = {37220997},
   publisher = {John Wiley and Sons Ltd},
   title = {Adaptively leveraging external data with robust meta-analytical-predictive prior using empirical Bayes},
   volume = {22},
   year = {2023},
}

@article{Pocock1976,
   author = {Stuart J. Pocock},
   doi = {10.1016/0021-9681(76)90044-8},
   issn = {00219681},
   issue = {3},
   journal = {Journal of Chronic Diseases},
   month = {3},
   pages = {175-188},
   title = {The combination of randomized and historical controls in clinical trials},
   volume = {29},
   year = {1976},
}

@article{Roychoudhury2020,
   author = {Satrajit Roychoudhury and Beat Neuenschwander},
   doi = {10.1002/sim.8456},
   issn = {10970258},
   issue = {7},
   journal = {Statistics in Medicine},
   month = {3},
   pages = {984-995},
   pmid = {31985077},
   publisher = {John Wiley and Sons Ltd},
   title = {Bayesian leveraging of historical control data for a clinical trial with time-to-event endpoint},
   volume = {39},
   year = {2020},
}

@article{Saint-Hilary2019,
   author = {Gaelle Saint-Hilary and Valentine Barboux and Matthieu Pannaux and Mauro Gasparini and Veronique Robert and Gianluca Mastrantonio},
   doi = {10.1002/sim.8060},
   issn = {10970258},
   issue = {10},
   journal = {Statistics in Medicine},
   month = {5},
   pages = {1753-1774},
   pmid = {30548627},
   publisher = {John Wiley and Sons Ltd},
   title = {Predictive probability of success using surrogate endpoints},
   volume = {38},
   year = {2019},
}

@article{Callegaro2023.bis,
   author = {Andrea Callegaro and Naveen Karkada and Emmanuel Aris and Toufik Zahaf},
   doi = {10.1002/pst.2283},
   issn = {1539-1604},
   issue = {3},
   journal = {Pharmaceutical Statistics},
   month = {5},
   pages = {475-491},
   title = {Vaccine clinical trials with dynamic borrowing of historical controls: Two retrospective studies},
   volume = {22},
   year = {2023}
}

@article{dunoyer_accelerating_2011,
	title = {Accelerating access to treatments for rare diseases},
	volume = {10},
	copyright = {http://www.springer.com/tdm},
	issn = {1474-1776, 1474-1784},
	url = {https://www.nature.com/articles/nrd3493},
	doi = {10.1038/nrd3493},
	language = {en},
	number = {7},
	urldate = {2025-05-08},
	journal = {Nature Reviews Drug Discovery},
	author = {Dunoyer, Marc},
	month = jul,
	year = {2011},
	pages = {475--476},
}

@article{dunne_extrapolation_2011-1,
	title = {Extrapolation of {Adult} {Data} and {Other} {Data} in {Pediatric} {Drug}-{Development} {Programs}},
	volume = {128},
	issn = {0031-4005, 1098-4275},
	url = {https://publications.aap.org/pediatrics/article/128/5/e1242/30987/Extrapolation-of-Adult-Data-and-Other-Data-in},
	doi = {10.1542/peds.2010-3487},
	language = {en},
	number = {5},
	urldate = {2025-05-08},
	journal = {Pediatrics},
	author = {Dunne, Julia and Rodriguez, William J. and Murphy, M. Dianne and Beasley, B. Nhi and Burckart, Gilbert J. and Filie, Jane D. and Lewis, Linda L. and Sachs, Hari C. and Sheridan, Philip H. and Starke, Peter and Yao, Lynne P.},
	month = nov,
	year = {2011},
	pages = {e1242--e1249},
}

@article{schoenfeld_bayesian_2009,
	title = {Bayesian design using adult data to augment pediatric trials},
	volume = {6},
	copyright = {https://journals.sagepub.com/page/policies/text-and-data-mining-license},
	issn = {1740-7745, 1740-7753},
	url = {https://journals.sagepub.com/doi/10.1177/1740774509339238},
	doi = {10.1177/1740774509339238},
	language = {en},
	number = {4},
	urldate = {2025-05-08},
	journal = {Clinical Trials},
	author = {Schoenfeld, David A and {Hui Zheng} and Finkelstein, Dianne M},
	month = aug,
	year = {2009},
	pages = {297--304},
}

@article{Morita2008,
  author    = {Morita, Satoshi and Thall, Peter F. and Müller, Peter},
  title     = {Determining the Effective Sample Size of a Parametric Prior},
  journal   = {Biometrics},
  year      = {2008},
  volume    = {64},
  number    = {2},
  pages     = {595–602},
  doi       = {10.1111/j.1541-0420.2007.00888.x},
  url       = {https://doi.org/10.1111/j.1541-0420.2007.00888.x}
}

@article{Kleyner1996,
  author    = {Kleyner, Andre and Bhagath, Shrikar and Gasparini, Mauro and Robinson, Jeffrey
and Bender, Mark },
  title     = {Bayesian techniques to reduce the sample size in automotive electronics 
attribute testing},
  journal   = {Microelectronics and Reliability},
  year      = {1997},
  volume    = {37},
  number    = {6},
  pages     = {879-883},
 }

\newpage

\section*{Supplementary Material}

\subsection*{Proof of Theorem 1}

Consider a RCT where mean control and treatment responses are normal $X_c \sim \mathcal{N} \left(\theta_c, \sigma^2_c \right)$, $X_t \sim \mathcal{N} \left(\theta_t, \sigma^2_t \right)$, and assume $\sigma^2_t = K\sigma^2_c$ (where $K^{-1}$ is the randomization ratio, assumed > 1). Assume a RMP  $\pi_c(\theta_c) = \omega \pi_\text{inf}(\theta_c)+(1-\omega) \pi_\text{rob}(\theta_c)$ is used for the control parameter, where $\pi_\text{inf}(\theta_c)$ and $\pi_\text{rob}(\theta_c)$ are the PDF of normally distributed random variables with parameters $\mu_{\text{inf}}$, $\sigma^2_{\text{inf}}$ and $\mu_{\text{rob}}$, $\sigma^2_{\text{rob}}$ respectively; while a normal prior distribution $\theta_t \sim \mathcal{N} \left(\mu_t, \sigma^2_{\text{rob}} \right)$ is given to the treatment parameter. Consider the type I error rate $\alpha \left( \cdot \right)$ as defined in Equation (\ref{t1e}), corresponding to the null hypothesis $H_0: \theta_c = \theta_t = D + \mu_\text{inf}$, where $D=\theta_c-\mu_{\text{inf}}$ is the drift parameter.
Then the following hold:
\begin{equation*}
    \lim_{D \rightarrow + \infty} \alpha \left( D + \mu_{\text{inf}}\right) = \eta \; \; \; \Longleftrightarrow \; \; \;  \lim_{D \rightarrow +\infty} \frac{D}{\sigma^2_{\text{rob}}} = 0
\end{equation*}

\begin{proof} Consider the following change of variable: $H=D+\mu_{\text{inf}}$, so that the thesis of the theorem becomes: 
\begin{equation*}
    \lim_{H \rightarrow + \infty} \alpha \left( H\right) = \eta \; \; \; \Longleftrightarrow \; \; \;  \lim_{H \rightarrow +\infty} \frac{H}{\sigma^2_{\text{rob}}} = 0 \; .
\end{equation*}
Since under the null hypotheses $\theta_c=\theta_t=H$  control and treatment responses are respectively $X_c \sim \mathcal{N} \left(H, \sigma^2_c \right)$ and $X_t \sim \mathcal{N} \left(H, \sigma^2_t \right)$, then the observed mean responses can be expressed as $X_c = H+\Delta_c$, where $\Delta_c \sim \mathcal{N} \left( 0, \sigma^2_c \right)$ and $X_t = H+ \Delta_t$, where $\Delta_t \sim \mathcal{N} \left( 0, \sigma^2_t \right)$. \\
\noindent
It follows from Equation (\ref{update normal}) that 
\begin{equation*}
    \lim_{H \rightarrow + \infty} \Tilde{\Omega} \left( X_c \right) = \lim_{H \rightarrow + \infty} \Tilde{\Omega} \left( H+\Delta_c \right) = \lim_{H \rightarrow + \infty} \Tilde{\Omega} \left( H \right) = 0  \; \; \Longrightarrow \; \;
    \lim_{H \rightarrow + \infty} \Tilde{\omega} \left( X_c \right) = 0
\end{equation*}
\noindent
where the second equality holds since $\Delta_c \sim o(H)$ for $H \rightarrow +\infty$. \\
As a consequence Equation (\ref{RMP update}) reduces to 
\begin{equation*}
    \lim_{H \rightarrow + \infty} g(\theta_c \; | \; x_c, \pi_{\text{inf}}, \pi_{\text{rob}}) = \lim_{H \rightarrow + \infty} g_{\text{rob}}(\theta_c | x_c, \pi_{\text{rob}})
\end{equation*}
\noindent
where $g_{\text{rob}}(\cdot | x_c, \pi_{\text{rob}})$ is the PDF of a normal distribution  $\mathcal{N} \left(\mu^{\text{post}}_{\text{c}},  \sigma^{2,\text{post}}_{\text{c}} \right)$, with 
\begin{equation*}
    \mu^{\text{post}}_{\text{c}} = \frac{\sigma^2_{\text{rob}}x_c + \sigma^2_c \mu_\text{rob}}{\sigma^2_c + \sigma^2_{\text{rob}}} = 
    \frac{\sigma^2_{\text{rob}}H + 
    \sigma^2_{\text{rob}}\Delta_c + \sigma^2_c \mu_\text{rob}}{\sigma^2_c + \sigma^2_{\text{rob}}}
    \; \; \; \; \; \; \; \;  \; \; \; \; \; \; \; 
\sigma^{2,\text{post}}_{\text{c}} = \frac{\sigma^2_c \sigma^2_{\text{rob}}}{\sigma^2_c + \sigma^2_{\text{rob}}} \tag{T1.1}\label{eq:T1.1}
\end{equation*}
\noindent
Using the same argument the posterior distribution for $\theta_t$ is $\mathcal{N} \left(\mu^{\text{post}}_{t},  \sigma^{2,\text{post}}_{t} \right)$; with 
\begin{equation*}
    \mu^{\text{post}}_{t} = \frac{\sigma^2_{\text{rob},t}x_t + K\sigma^2_c \mu_t}{K\sigma^2_c + \sigma^2_{\text{rob},t}} = 
    \frac{ \sigma^2_{\text{rob},t}H +\sigma^2_{\text{rob},t}\Delta_t + K\sigma^2_c \mu_t}{K\sigma^2_c + \sigma^2_{\text{rob},t}} \; \; \; \; \; \; \; \;  \; \; \; \; \; \; 
\sigma^{2,\text{post}}_{t} = \frac{K\sigma^2_c \sigma^2_{\text{rob},t}}{K\sigma^2_c + \sigma^2_{\text{rob},t}} \tag{T1.2}\label{eq:T1.2}
\end{equation*}
\noindent
Since the posterior densities for $\theta_c$ and $\theta_t$ are normally distributed, then the posterior probability for the mean treatment difference parameter is normal itself, i.e. $\delta^{\text{post}}  \sim \mathcal{N} \left( \mu^\text{post}_t - \mu^\text{post}_c, \sigma^{2,\text{post}}_t + \sigma^{2,\text{post}}_c \right)$. Notice that while the variance of the latter distribution is a fixed quantity, as it does not depend on $H$; the mean is a random variable depending on $\Delta_c$ and $\Delta_t$. \\
Let us prove the two implications of the Theorem separately.

\vspace{0.5 cm}
\noindent{ \fbox{$\Longrightarrow$} \hspace{0.5cm} Let us proceed by contradiction. If  $\lim_{H \rightarrow +\infty} \frac{H}{\sigma^2_{\text{rob}}} = +\infty$, then exploiting the equalities in \ref{eq:T1.1} and \ref{eq:T1.2}, and ignoring negligible terms it holds that:}
\begin{equation*}
    \lim_{H \rightarrow +\infty} \mu^\text{post}_t - \mu^\text{post}_c = \frac{H(1-K)\sigma^2_{\text{rob}}\sigma^2_c}{(K\sigma^2_c + \sigma^2_{\text{rob}})(\sigma^2_c + \sigma^2_{\text{rob}})} = +\infty \; \; \; \; \; \;   \forall x_c, x_t \in \mathbb{R}
\end{equation*}
and from Equation (\ref{success RCT}) follows that 
\begin{equation*}
    \lim_{H \rightarrow +\infty} \mathbb{P}\left( \delta > 0 \;|\; x_c, x_t \right) = \Phi\left(+\infty\right) = 1 > 1-\eta \; \; \; \; \forall x_c, x_t \in \mathbb{R}
\end{equation*}
meaning that success is achieved with probability 1 as $H \rightarrow +\infty$, and accordingly
\begin{equation*}
    \lim_{H \rightarrow +\infty} \vmathbb{1} \left\{ \mathbb{P}\left( \delta > 0 \;|\; x_c, x_t \right)\right\} = \vmathbb{1} \left\{ \left(-\infty, + \infty) \times (-\infty, + \infty) \right) \right\}
\end{equation*}
Type I error $\alpha(D+\mu_{\text{inf}})$ is easily obtained by integrating the success over the  likelihood
\begin{equation*}
\begin{split}
    \lim_{H \rightarrow + \infty} \alpha \left( H \right)  = & \lim_{H \rightarrow + \infty} \iint _{\mathbb{R}^2} \vmathbb{1} \left\{  \mathbb{P}\left( \delta > 0 \;|\; x_c, x_t \right) > \eta \right\} f_{X_c}(x_c|\theta_c=H) f_{X_t}(x_t|\theta_t=H) \;  dx_c \;dx_t \\
       = & \iint _{\mathbb{R}^2} \lim_{H \rightarrow + \infty} \vmathbb{1}  \left\{  \mathbb{P}\left( \delta > 0 \;|\; x_c, x_t \right) > \eta \right\} f_{X_c}(x_c|\theta_c=H) f_{X_t}(x_t|\theta_t=H) \;  dx_c \;dx_t \\
    = & \iint _{\mathbb{R}^2} f_{X_c}(x_c|\theta_c=H) f_{X_t}(x_t|\theta_t=H) \;  dx_c \;dx_t = 1  
\end{split}   
\end{equation*} 

\vspace{0.5 cm}

 \noindent{\fbox{$\Longleftarrow$} \hspace{0.5 cm} If  $\lim_{H \rightarrow +\infty} \frac{H}{\sigma^2_{\text{rob}}} \neq +\infty$, then exploiting the equalities in \ref{eq:T1.1} and \ref{eq:T1.2}, and ignoring negligible terms it holds that:}
\begin{equation*}
    \lim_{H \rightarrow +\infty} \mu^\text{post}_t - \mu^\text{post}_c = x_t - x_c \; \; \; \; \; \; \; \; \; \;
    \lim_{H \rightarrow +\infty} \sigma^{2,\text{post}}_c = \sigma^2_c \; \; \; \; \; \; \; \;\; \;
    \lim_{H \rightarrow +\infty} \sigma^{2,\text{post}}_t = \sigma^2_t
\end{equation*}
\noindent
and from Equation (\ref{success RCT}) follows that
\begin{equation*}
    \lim_{H \rightarrow +\infty} \mathbb{P}\left( \delta > 0 \;|\; x_c, x_t \right)  > 1-\eta \; \; \Longleftrightarrow \frac{x_t - x_c}{\sqrt{\sigma_t^2 + \sigma_c^2}} > z_{\eta}
\end{equation*}
where $z_{\eta}$ is the $\eta$ quantile of a standard normal distribution. \\
The limit of the type I error for $H \rightarrow +\infty$ is:
\begin{equation*}
\begin{split}
    \lim_{H \rightarrow + \infty} \alpha \left( H \right)  = & \lim_{H \rightarrow + \infty} \iint _{\mathbb{R}^2} \vmathbb{1} \left\{  \mathbb{P}\left( \delta > 0 \;|\; x_c, x_t \right) > \eta \right\} f_{X_c}(x_c|\theta_c=H) f_{X_t}(x_t|\theta_t=H) \;  dx_c \;dx_t \\
       = & \iint _{\mathbb{R}^2} \vmathbb{1}  \left\{  \mathbb{P}\left( \delta > 0 \;|\; x_c, x_t \right) > \eta \right\} f_{X_c}(x_c|\theta_c=H) f_{X_t}(x_t|\theta_t=H) \;  dx_c \;dx_t \\
    = & \iint _{\mathbb{R}^2} \vmathbb{1}  \left\{ \frac{x_t - x_c}{\sqrt{\sigma_t^2 + \sigma_c^2}} > z_{\eta} \right\} f_{X_c}(x_c|\theta_c=H) f_{X_t}(x_t|\theta_t=H) \;  dx_c \;dx_t \\
    = & \int_{z_{\eta}\sqrt{\sigma_t^2 + \sigma_c^2}}^{+\infty} f_{X_t-X_c}(\xi) d \xi = 1 - \Phi \left( z_{\eta} \right) = \eta
\end{split}   
\end{equation*} 
\noindent
where $\xi=x_t - x_c$ and the last equality follows from the fact that $X_t - X_c \sim \mathcal{N} \left( 0, \sigma_t^2 + \sigma_c^2 \right)$
\end{proof}

\subsection*{Proof of Theorem 2}
Consider a normal random variable modeling the mean control response $X_c \sim \mathcal{N} \left(\theta_c, \sigma^2_c \right)$, and assume two distinct RMPs are used for the underlying parameter $\theta_c$, namely
\begin{equation*}
    \pi^{(1)}_c(\theta_c) = \omega \pi_\text{inf}(\theta_c)+(1-\omega) \pi_\text{rob}^{(1)}(\theta_c) \; \; \; \; \; \; \; \; \pi^{(2)}_c(\theta_c) = \omega \pi_\text{inf}(\theta_c)+(1-\omega) \pi_\text{rob}^{(2)}(\theta_c)
\end{equation*} 
where $\pi_\text{inf}(\theta_c)$ and $\pi^{(i)}_\text{rob}(\theta_c)$ are the PDF of normally distributed random variables with parameters $\mu_{\text{inf}}$, $\sigma^2_{\text{inf}}$ and $\mu^{(i)}_{\text{rob}}$, $\sigma^2_{\text{rob}}$ respectively with $i \in \{ 1,2 \}$. \\
Consider the posterior distributions $ g(\theta_c  |  x_c, \pi^{(1)}_c)$ and $ g(\theta_c  |  x_c, \pi^{(2)}_c)$, then
\begin{equation*}
    \lim_{\sigma^2_{\text{rob}} \rightarrow + \infty}  g(\theta_c  |  x_c, \pi^{(1)}_c) = \lim_{\sigma^2_{\text{rob}} \rightarrow + \infty} g(\theta_c  |  x_c, \pi^{(2)}_c) \; \; \; \; \; \; \; \; \; \; \; \; \; \; \forall x_c \in \mathbb{R}
\end{equation*}

\begin{proof}
    The two RMPs for $\theta_c$ differ only for the the locations of their robustification components, which impact the posterior weights $\Tilde{\omega}$ and the posterior corresponding to the robustification component $g_{\text{rob}}(\theta_c | x_c, \pi^{(i)}_{\text{rob}})$. In the following, the argument will be proven by working independently on these two objects. \\
    Given Equation (\ref{Odds update}), it holds that for $\sigma^2_{\text{rob}} \rightarrow +\infty$, then
    \begin{equation*}
        \frac{1}{R^2}{\frac{\left( x_c -\mu_{\text{rob}} \right)^2}{2v_\text{inf}^2}} \sim o \left( \frac{d^2}{2v_\text{inf}^2} \right) \; \; \; \; \; \; \; \Longrightarrow \; \; \; \; \; \; \;
        \Tilde{\Omega} \sim \frac{\Omega}{R} \exp\left\{\frac{d^2}{2v_\text{inf}^2}\right\}\tag{T2.1}\label{eq:T2.1} \; .
    \end{equation*}
    \noindent
    The latter is independent on $\mu^{(i)}_{\text{rob}}$; as a consequence
    \begin{equation*}
        \lim_{\sigma^2_{\text{rob}} \rightarrow +\infty}\Tilde{\omega}(x_c; \pi_{\text{inf}}, \pi^{(1)}_{\text{rob}}, \omega) = \lim_{\sigma^2_{\text{rob}}\rightarrow +\infty}\Tilde{\omega}(x_c; \pi_{\text{inf}}, \pi^{(2)}_{\text{rob}}, \omega) \; \; \; \; \; \; \; \; \; \; \; \; \; \; \forall x_c \in \mathbb{R}\tag{T2.2}\label{eq:T2.2}
    \end{equation*}
    \noindent
    Moreover, the posterior distribution $g_{\text{rob}}(\theta_c | x_c, \pi^{(i)}_{\text{rob}})$ corresponding to each robustification component is normal with parameters $\mu^{(i), \text{post}}_{\text{rob}}$ and $\sigma^{2, \text{post}}_{\text{rob}}$, with

    \begin{equation*}
        \mu^{(i), \text{post}}_{\text{rob}} = \frac{\sigma^2_{\text{rob}}x_c + \sigma^2_c \mu^{(i)}_\text{rob}}{\sigma^2_c + \sigma^2_{\text{rob}}} \; \; \; \; \; \; \; \; \; \; \; \; \; \; \; \; \sigma^{2,\text{post}}_{\text{c}} = \frac{\sigma^2_c \sigma^2_{\text{rob}}}{\sigma^2_c + \sigma^2_{\text{rob}}}
    \end{equation*}
\noindent
Notice that the variance, which is the same in the two RMPs, does not depend on $\mu^{(i)}_{\text{rob}}$, moreover for the mean we have that for $\sigma^2_{\text{rob}} \rightarrow +\infty$, then
\begin{equation*}
    \mu^{(i), \text{post}}_{\text{rob}} \sim \frac{\sigma^2_{\text{rob}}x_c}{\sigma^2_c + \sigma^2_{\text{rob}}}
\end{equation*}
\noindent
which is independent on $\mu^{(i)}_{\text{rob}}$. It follows that
\begin{equation*}
    \lim_{\sigma^2_{\text{rob}} \rightarrow +\infty}g_{\text{rob}}(\theta_c | x_c, \pi^{(1)}_{\text{rob}}) = \lim_{\sigma^2_{\text{rob}} \rightarrow +\infty}g_{\text{rob}}(\theta_c | x_c, \pi^{(2)}_{\text{rob}}) \tag{T2.3}\label{eq:T2.3}
\end{equation*}
\noindent
The argument follows from Equation (\ref{eq:T2.2}) and \ref{eq:T2.3}.
\end{proof}

\subsection*{Proof of Theorem 3}

    Consider a normal random variable $X_c \sim \mathcal{N} \left(\theta_c, \sigma^2_c \right)$, and assume a RMP is used for the parameter $\theta_c$, namely $\pi_c(\theta_c) = \omega \pi_\text{inf}(\theta_c)+(1-\omega) \pi_\text{rob}(\theta_c)$, where $\pi_\text{inf}(\theta_c)$ and $\pi_\text{rob}(\theta_c)$ are the PDF of normally distributed random variables with parameters $\mu_{\text{inf}}$, $\sigma^2_{\text{inf}}$ and $\mu_{\text{rob}}$, $\sigma^2_{\text{rob}}$ respectively. The following hold:
    \begin{enumerate}
        \item if $\Omega < + \infty$, then
        \begin{equation*}
        \lim_{\sigma^2_{\text{rob}} \rightarrow + \infty} \Tilde{\omega} \left(x_c, \pi_\text{inf}(\theta_c), \pi_\text{rob}(\theta_c), \omega \right) = 1 \; \; \; \; \; \; \; \forall x_c \in \left( -\infty, +\infty \right)
        \end{equation*}

        \begin{proof}
        From the asymptotic equivalence in \ref{eq:T2.1}, considering that $\Omega<+\infty$ and considering that $R \rightarrow +\infty$ for $\sigma^2_{\text{rob}} \rightarrow + \infty$, then the argument follows.
        \end{proof}
        
        \item if $\Omega \sim O(R)$ for $\sigma^2_{\text{rob}} \rightarrow + \infty$, then
        \begin{equation*}
            \lim_{\sigma^2_{\text{rob}} \rightarrow + \infty} \Tilde{\omega} \left(x_c, \pi_\text{inf}(\theta_c), \pi_\text{rob}(\theta_c), \omega \right) \neq 1 \; \; \; \; \; \; \; \forall x_c \in \left( -\infty, +\infty \right)
        \end{equation*}

        \begin{proof}
        From the asymptotic equivalence in \ref{eq:T2.1}, considering that $\Omega \sim O(R) \Rightarrow  \beta \left(\omega,R \right)<+\infty$ for $\sigma^2_{\text{rob}} \rightarrow + \infty$, then the argument follows. 
        \end{proof}       
    \end{enumerate}

\subsection*{Proof of Theorem 4}

Consider a binomial random variable $X_c \sim \text{Bin} \left(\theta_c, n_c \right)$, and assume a RMP is used for the parameter $\theta_c$, namely $\pi_c(\theta_c) = \omega \pi_\text{inf}(\theta_c)+(1-\omega) \pi_\text{rob}(\theta_c)$, where $\pi_\text{inf}(\theta_c)$ and $\pi_\text{rob}(\theta_c)$ are the PDF of Beta distributed random variables with parameters $a_{\text{inf}}$, $b_{\text{inf}}$ and $a_{\text{rob}} = b_{\text{rob}} = \varepsilon$, respectively. The following hold:
    \begin{enumerate}
        \item if $\Omega < + \infty$, then
        \begin{equation*}
        \lim_{\varepsilon \rightarrow 0} \Tilde{\omega} \left(x_c, \pi_\text{inf}(\theta_c), \pi_\text{rob}(\theta_c), \omega \right) = 1 \; \; \; \; \; \; \; \forall x_c \in \left( 0, n_c \right)
        \end{equation*}

        \begin{proof}
            From Equation \eqref{beta_fun_betabinom}, and expressing the Beta function using the Gamma functions $B(x,y)=\Gamma(a) \Gamma (b) / \Gamma(a+b)$, the posterior odds under the Robust Mixture Prior (RMP) in the Beta-Binomial setting can be written as
\begin{equation*}
\begin{split}
    \Omega(x_c) = \beta(\omega, a_\text{rob}, b_\text{rob}) 
& \times \frac{\Gamma(x_c + a_\text{inf}) \Gamma(n_c - x_c + b_\text{inf}) \Gamma(a_\text{inf} + b_\text{inf})}
{\Gamma(n_c + a_\text{inf} + b_\text{inf}) \Gamma(a_\text{inf}) \Gamma(b_\text{inf})} \\
& \times 
\frac{\Gamma(n_c + a_\text{rob} + b_\text{rob})}
{\Gamma(x_c + a_\text{rob}) \Gamma(n_c - x_c + b_\text{rob})}
,
\end{split}
\end{equation*}

where
\begin{equation*}
    \beta(\omega, a_\text{rob}, b_\text{rob}) = 
\frac{\omega}{1-\omega} 
\cdot
\frac{\Gamma(a_\text{rob}) \Gamma(b_\text{rob})}
{\Gamma(a_\text{rob} + b_\text{rob})}.
\end{equation*}

\noindent
Under the assumptions of the theorem 
\(a_\text{rob} = b_\text{rob} = \varepsilon\) with \(\varepsilon \to 0^+\), and using the well-known asymptotic expansion \(\Gamma(\varepsilon) \sim 1/\varepsilon\) as \(\varepsilon \to 0^+\), 
and the fact that \(\Gamma(x_c + \varepsilon) \to \Gamma(x_c)\) for \(x_c > 0\), we obtain
\[
\Gamma(a_\text{rob}) \Gamma(b_\text{rob}) \sim \frac{1}{\varepsilon^2}, 
\quad
\Gamma(a_\text{rob} + b_\text{rob}) = \Gamma(2\varepsilon) \sim \frac{1}{2\varepsilon},
\]
and \(\Gamma(n_c + a_\text{rob} + b_\text{rob}) \sim \Gamma(n_c)\).

Substituting these limits into the definition of \(\beta(\omega, a_\text{rob}, b_\text{rob})\) gives
\begin{equation*}
\beta(\omega, a_\text{rob}, b_\text{rob}) 
\sim 
\frac{\omega}{1-\omega} \cdot \frac{2}{\varepsilon} \to +\infty \quad \text{as} \quad \varepsilon \to 0
\end{equation*}

\noindent
The remaining multiplicative factor in the expression for \(\Tilde{\Omega}(x_c)\),
\begin{equation*}
    C(x_c, n_c) = 
\frac{B \left( a_{\text{inf}} + x_c, b_{\text{inf}} + n_c - x_c\right)}{B \left( x_c, n_c - x_c\right) B \left(a_{\text{inf}}, b_{\text{inf}} \right)} \; ,
\end{equation*}

\noindent
is finite and positive for all \(x_c \in (0,n_c)\). Therefore,
\begin{equation*}
\Tilde{\Omega}(x_c) 
= \beta(\omega, a_\text{rob}, b_\text{rob}) \cdot C(x_c, n_c)
\to +\infty \quad \text{as } \varepsilon \to 0^+.
\end{equation*}

\noindent
Finally, the posterior weight of the informative component is
\[
\Tilde{\omega}\left(x_c, \pi_\text{inf}(\theta_c), \pi_\text{rob}(\theta_c), \omega \right)
= \frac{\Tilde{\Omega}(x_c)}{1 + \Tilde{\Omega}(x_c)}.
\]
Since \(\Tilde{\Omega}(x_c) \to +\infty\), it follows that
\[
\lim_{\varepsilon \to 0^+} 
\Tilde{\omega}\left(x_c, \pi_\text{inf}(\theta_c), \pi_\text{rob}(\theta_c), \omega \right) = 1,
\quad \forall x_c \in (0,n_c).
\]

\end{proof}
        
        \item if $\Omega \sim O (\varepsilon)$ for $ \varepsilon \rightarrow 0$, then
        \begin{equation*}
            \lim_{\varepsilon \rightarrow 0} \Tilde{\omega} \left(x_c, \pi_\text{inf}(\theta_c), \pi_\text{rob}(\theta_c), \omega \right) \neq 1 \; \; \; \; \; \; \; \forall x_c \in \left( 0, n_c \right)
        \end{equation*} 
        
        \begin{proof}

Assume again that \(a_\text{rob} = b_\text{rob} = \varepsilon\) with \(\varepsilon \to 0^+\).  
In Point~1, we observed that as \(\varepsilon \to 0^+\), 
\(\Gamma(\varepsilon) \sim 1/\varepsilon\) and \(\Gamma(2\varepsilon) \sim 1/(2\varepsilon)\), 
so that \(\beta(\omega, \varepsilon, \varepsilon)\) diverges as \(O(1/\varepsilon)\).  
This divergence was responsible for \(\Omega(x_c) \to +\infty\), leading to \(\Tilde{\omega} \to 1\).

Here, we relax the assumption of a fixed \(\omega\) and instead assume that \(\Omega(x_c)\) satisfies the asymptotic scaling
\[
\Omega \sim O\!\left(\varepsilon \right) \quad \text{as } \varepsilon \to 0^+,
\]

This means that \(\Omega(x_c)\) and \(\varepsilon\) are of the same order of magnitude, i.e.
\[
\frac{\Omega}{\varepsilon} \to K,
\]
for some finite, positive constant \(K>0\).

\noindent
It follows that as $\varepsilon \to 0^+$,

\begin{equation*}
\begin{split}
    \Tilde{\Omega}(x_c) 
& = \beta(\omega, \varepsilon, \varepsilon) \cdot C(x_c, n_c) \\
& = K \cdot C(x_c, n_c) = \Tilde{K} < +\infty
\end{split}
\end{equation*}

\noindent
Substituting this asymptotic behavior into the expression for the posterior weight,
\[
\Tilde{\omega}\left(x_c, \pi_\text{inf}(\theta_c), \pi_\text{rob}(\theta_c), \omega \right) = 
\frac{\Tilde{\Omega}(x_c)}{1 + \Tilde{\Omega}(x_c)},
\]
we obtain that as \(\varepsilon \to 0^+\),
\[
\lim_{\varepsilon \to 0^+} \Tilde{\omega}\left(x_c, \pi_\text{inf}(\theta_c), \pi_\text{rob}(\theta_c), \omega \right) 
= \frac{ \Tilde{K}}{1 + \Tilde{K}} < 1, \quad \forall x_c \in (0,n_c).
\]

        \end{proof}
    \end{enumerate}

\subsection*{Proof of Equations (\ref{RMP update}) and (\ref{Weights Update})}

\begin{equation*}
\begin{split}
    g\left(\theta_c | x_c, \pi_c \right) & = \frac{\big[\omega\pi_{\text{inf}}(\theta_c) + (1-\omega)\pi_\text{rob} (\theta_c)\big]f\left(x_c | \theta_c \right)}{\int_{-\infty}^{+\infty}\big[\omega\pi_{\text{inf}}(\theta_c) + (1-\omega)\pi_\text{rob} (\theta_c)\big]f\left(x_c | \theta_c \right) d\theta_c} = \\[12pt]
    & = \frac{\omega\pi_{\text{inf}}(\theta_c)f\left(x_c | \theta_c \right) + (1-\omega)\pi_\text{rob} (\theta_c)f\left(x_c | \theta_c \right)}{\omega \int_{-\infty}^{+\infty}\pi_{\text{inf}}(\theta_c)f\left(x_c | \theta_c \right) d\theta_c + (1-\omega) \int_{-\infty}^{+\infty} \pi_\text{rob} (\theta_c)f\left(x_c | \theta_c \right) d\theta_c} = \\[12pt]
    & = \frac{\omega\pi_{\text{inf}}(\theta_c)f\left(x_c | \theta_c \right) + (1-\omega)\pi_\text{rob} (\theta_c)f\left(x_c | \theta_c \right)}{\omega f\left(x_c | \pi_{\text{inf}} \right) + (1-\omega) f\left(x_c | \pi_\text{rob} \right)} = \\[12pt]
    & = \frac{\omega\pi_{\text{inf}}(\theta_c)f\left(x_c | \theta_c \right)}{\omega f\left(x_c | \pi_{\text{inf}} \right) + (1-\omega) f\left(x_c | \pi_\text{rob} \right)} + \frac{(1-\omega)\pi_\text{rob} (\theta_c)f\left(x_c | \theta_c \right)}{\omega f\left(x_c | \pi_{\text{inf}} \right) + (1-\omega) f\left(x_c | \pi_\text{rob} \right)} = \\[12pt]
     & =  \frac{f\left(x_c | \theta_c \right) \pi_{\text{inf}}\left(\theta_c \right)} {f\left(x_c | \pi_\text{inf}\right)} \times \frac{\omega f\left(x_c | \pi_{\text{inf}} \right)}{\omega f\left(x_c | \pi_{\text{inf}} \right) +  (1-\omega) f\left(x_c | \pi_\text{rob} \right)} \; + \\[12pt]
      & + \frac{f\left(x_c | \theta_c \right) \pi_{\text{rob}}\left(\theta_c \right)}{f\left(x_c | \pi_\text{rob} \right)} \times \frac{(1-\omega) f\left(x_c | \pi_\text{rob} \right)}{\omega f\left(x_c | \pi_{\text{inf}} \right) + (1-\omega) f\left(x_c | \pi_\text{rob} \right)} \;.
\end{split}
\end{equation*}

\subsection*{Formulas for the metrics used in posterior inference}
Bias is defined as:
\begin{equation*}
    b(\hat{\delta}) = \mathbb{E} \left[ \hat{\delta} - \delta  \right] = \iint _{\mathbb{R}^2} \left( \hat{\delta} - \delta \right) f_{X_c}(x_c) f_{X_t}(x_t) \, dx_c\,dx_t \;,
\end{equation*}
Variance is defined as:  
\begin{equation*}
    Var(\hat{\delta}) = \mathbb{E} \left[ \left(\hat{\delta} - \mathbb{E}\left[\delta\right] \right)^2  \right] = \iint _{\mathbb{R}^2} \left( \hat{\delta} - \mathbb{E}\left[\delta\right] \right)^2 f_{X_c}(x_c) f_{X_t}(x_t) \, dx_c\,dx_t
\end{equation*}
Mean Squared Error (MSE) is defined as:
\begin{equation*}
    MSE(\hat{\delta}) = \mathbb{E} \left[ \left(\hat{\delta} - \delta \right)^2  \right] = \iint _{\mathbb{R}^2} \left( \hat{\delta} - \delta \right)^2 f_{X_c}(x_c) f_{X_t}(x_t) \, dx_c\,dx_t
\end{equation*}

\newpage
\section*{Supplementary Figures}

\renewcommand{\thefigure}{S1}
\begin{figure}[H]
	\begin{subfigure}[$\omega=0.5$]{
		\centering
		\includegraphics[width=0.6\textwidth]{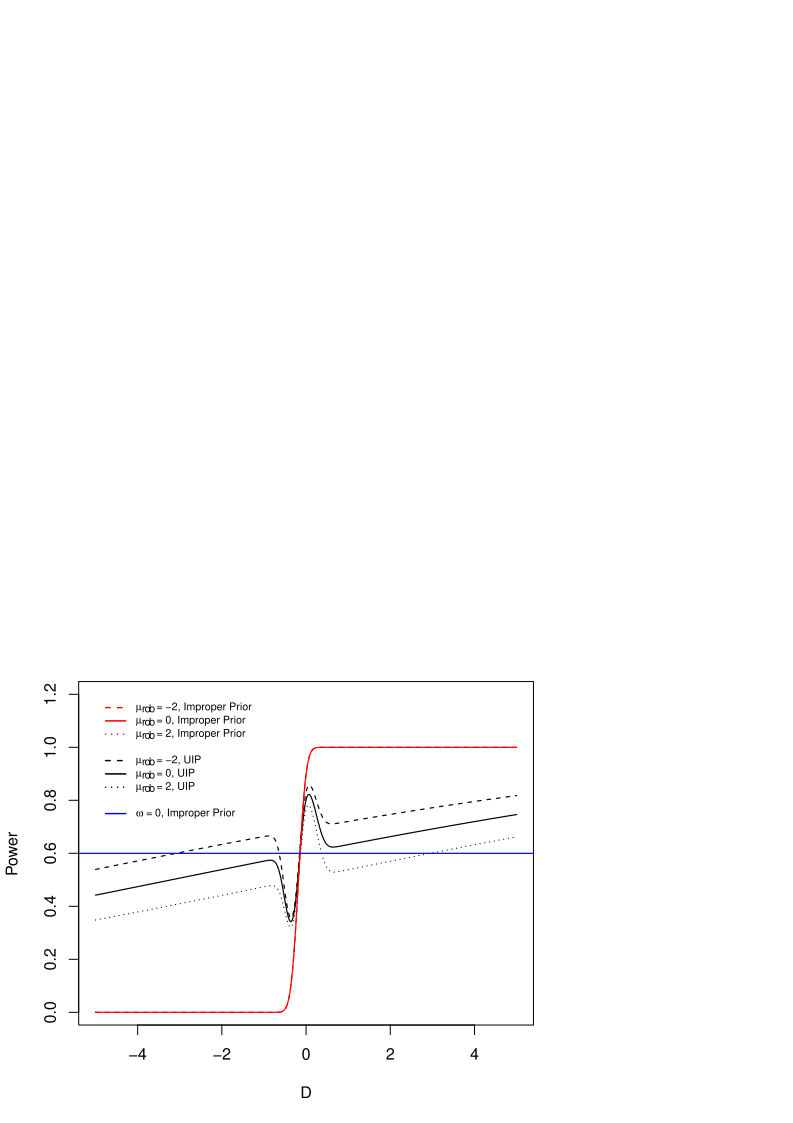}
        \label{pow.05.fig}}
	\end{subfigure} \hfill
	\begin{subfigure}[$\omega=0.9$]{
		\centering
		\includegraphics[width=0.6\textwidth]{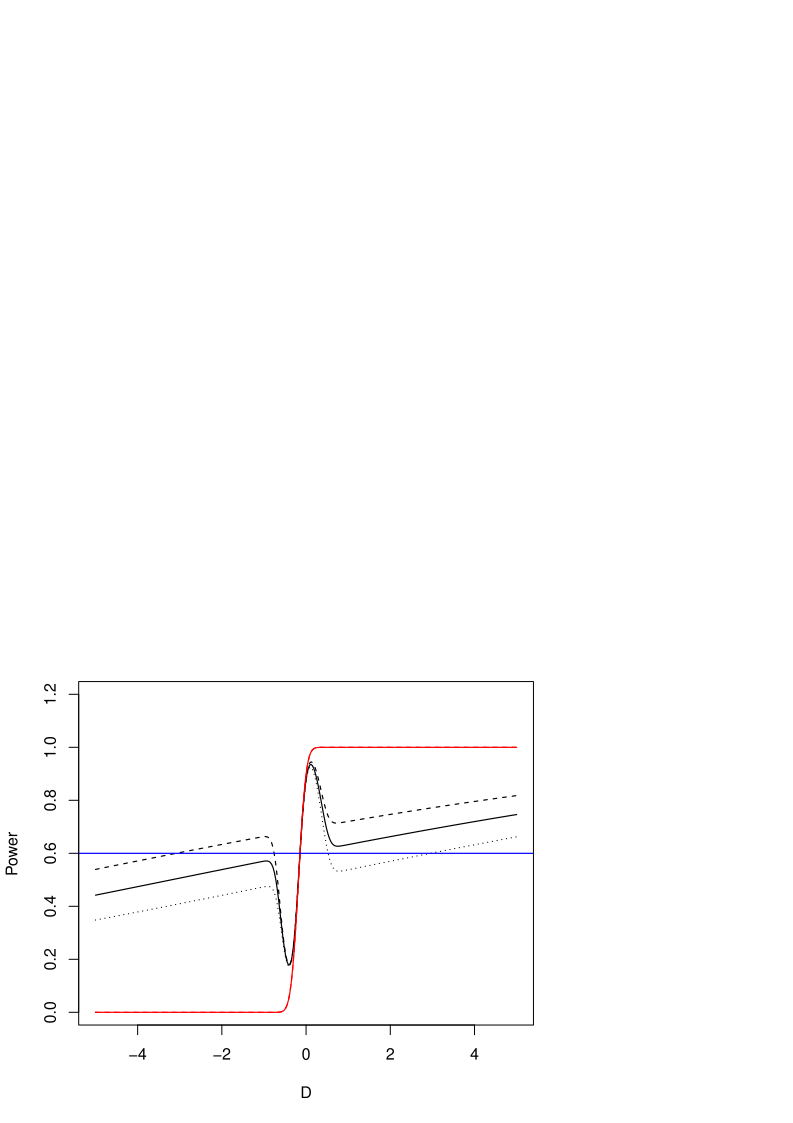}
        \label{pow.09.fig}}
	\end{subfigure}
	\caption{Power $\text{Pow}(D)$ under different choices of parameters for the RMP. Red curves: improper prior distributions ($\sigma^2_{\text{rob}} = 10^{100}$). Black curves: unit-information prior ($\sigma^2_{\text{rob}} = 1$). Different choices of $\mu_{\text{rob}}$ are denoted with different line types. Panel (a): analysis with prior mixture weight $\omega=0.5$. Panel (b): analysis with prior mixture weight $\omega=0.9$.}
    \label{motivating.exe.bis.fig}
\end{figure}

\renewcommand{\thefigure}{S2}
\begin{figure}[h!]
	\begin{subfigure}[$\beta^*=0.65$]{
		\centering
		\includegraphics[width=0.27\textwidth]{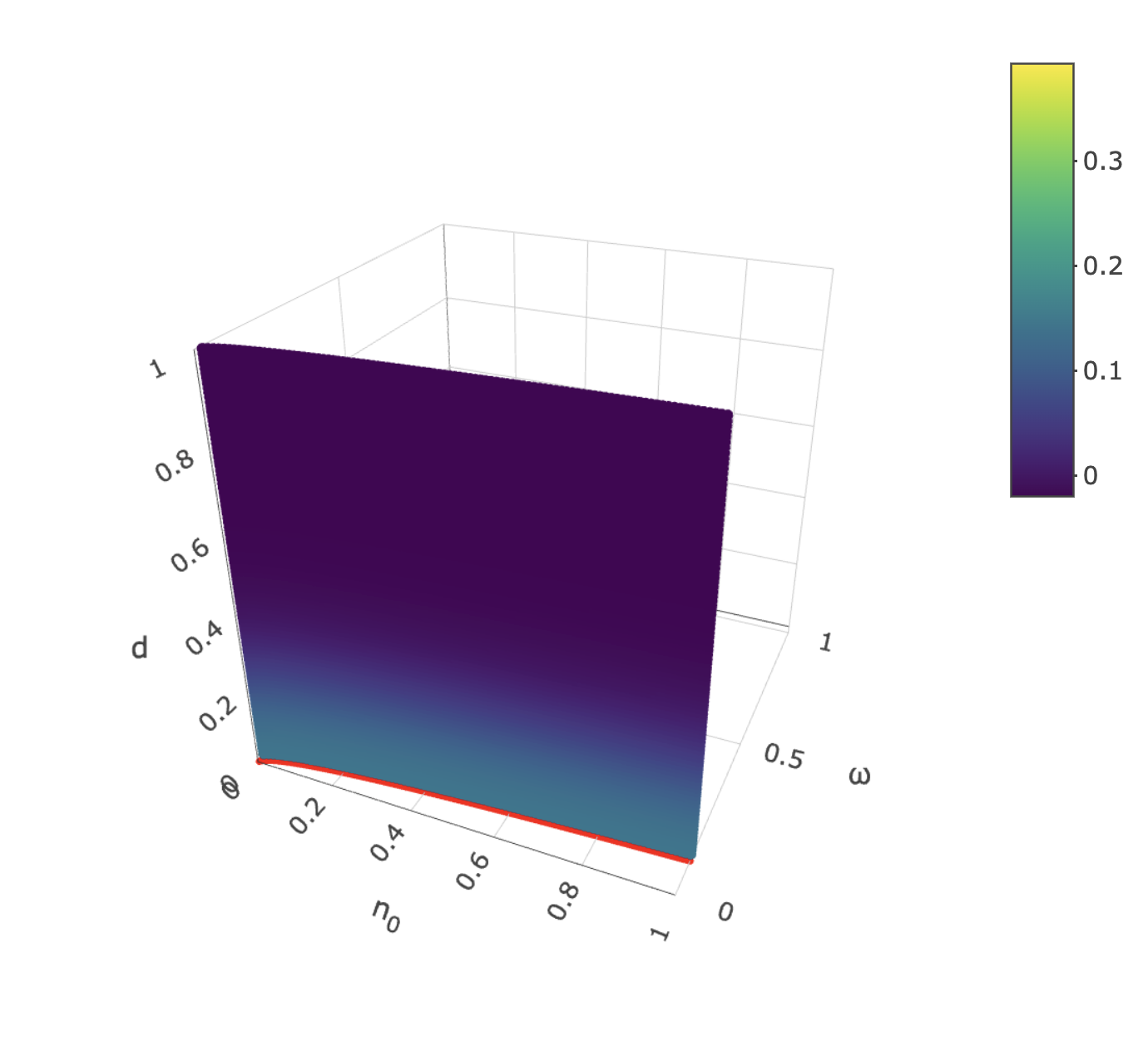}
        \label{p1}}
	\end{subfigure} \hfill
	\begin{subfigure}[$\beta^*=1.46$]{
		\centering
		\includegraphics[width=0.27\textwidth]{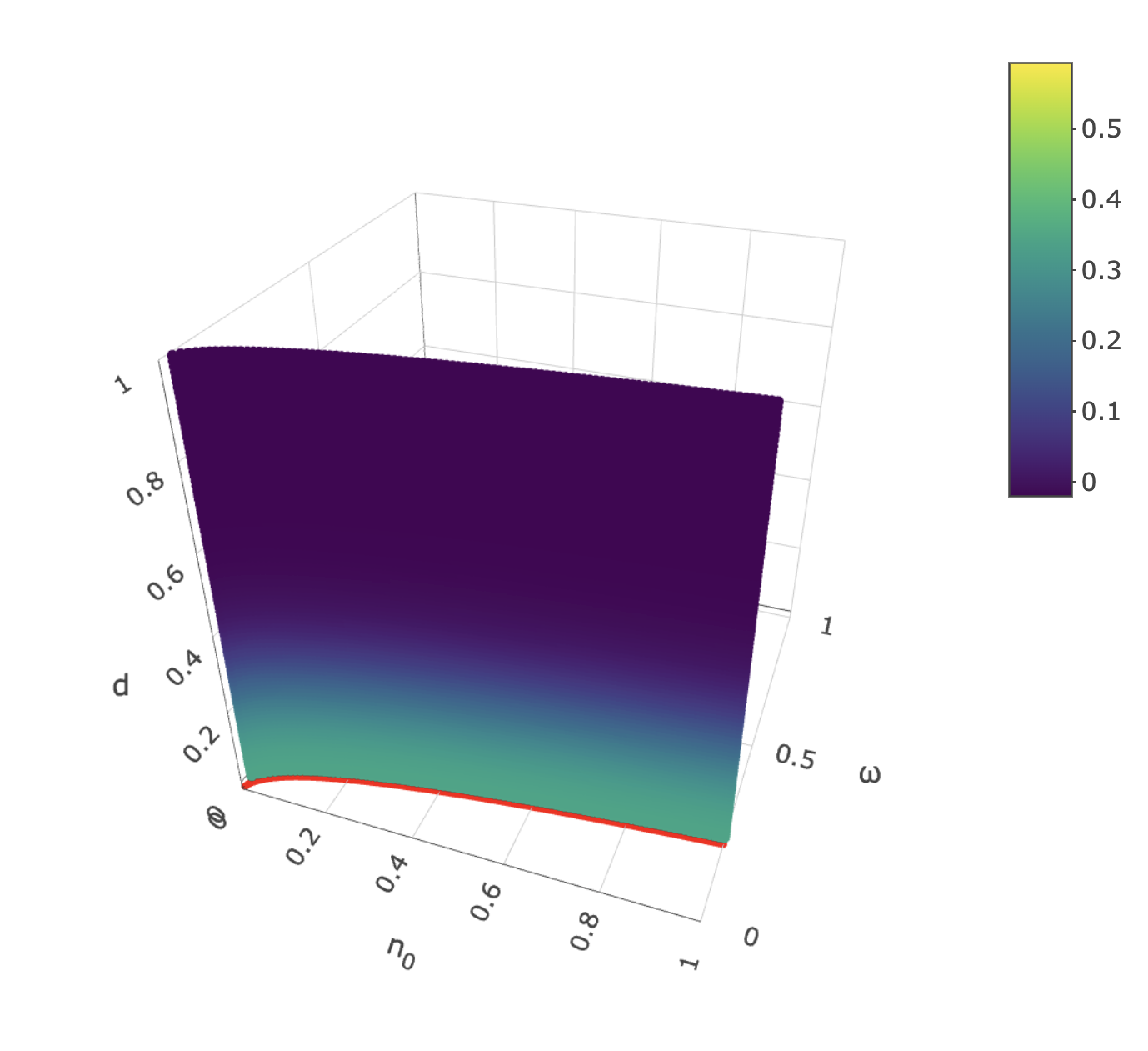}
        \label{p2}}
	\end{subfigure}\hfill
    \begin{subfigure}[$\beta^*=2.50$]{
		\centering
		\includegraphics[width=0.27\textwidth]{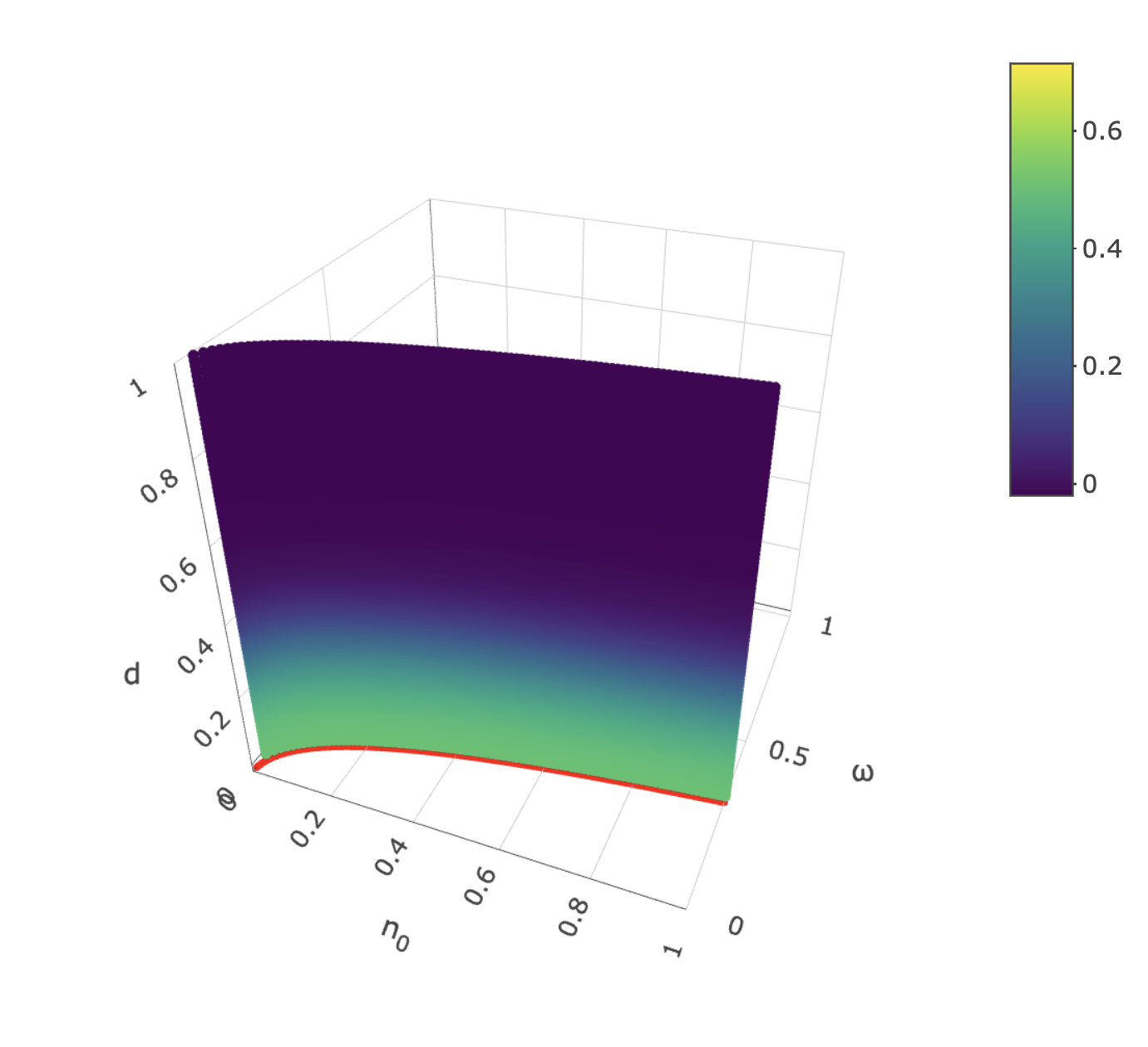}
        \label{p3}}
	\end{subfigure}
    \begin{subfigure}[$\beta^*=3.89$]{
		\centering
		\includegraphics[width=0.27\textwidth]{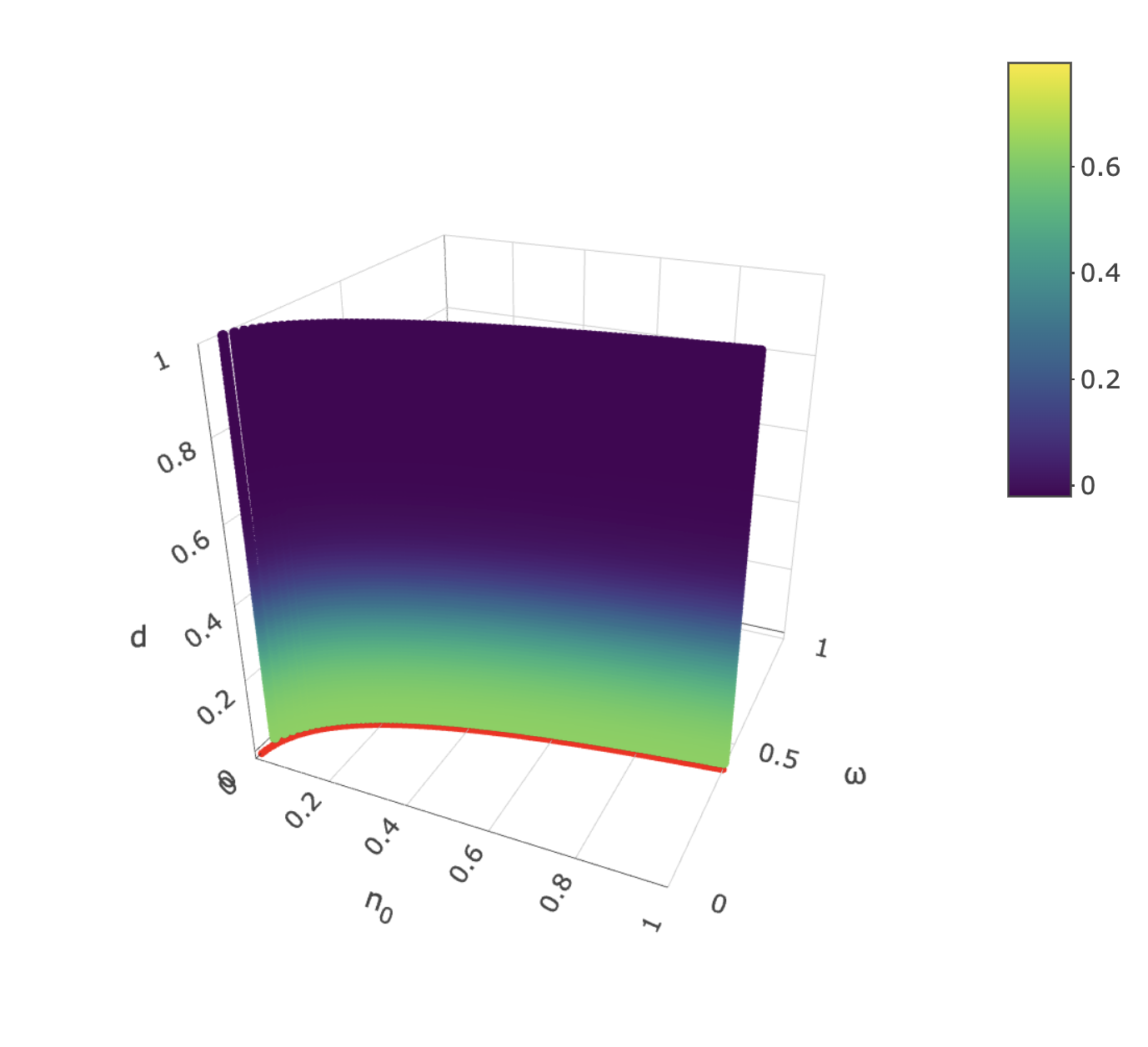}
        \label{p4}}
	\end{subfigure} \hfill
    \begin{subfigure}[$\beta^*=5.83$]{
		\centering
		\includegraphics[width=0.27\textwidth]{3D_05.png}
        \label{p5}}
	\end{subfigure} \hfill
	\begin{subfigure}[$\beta^*=8.75$]{
		\centering
		\includegraphics[width=0.27\textwidth]{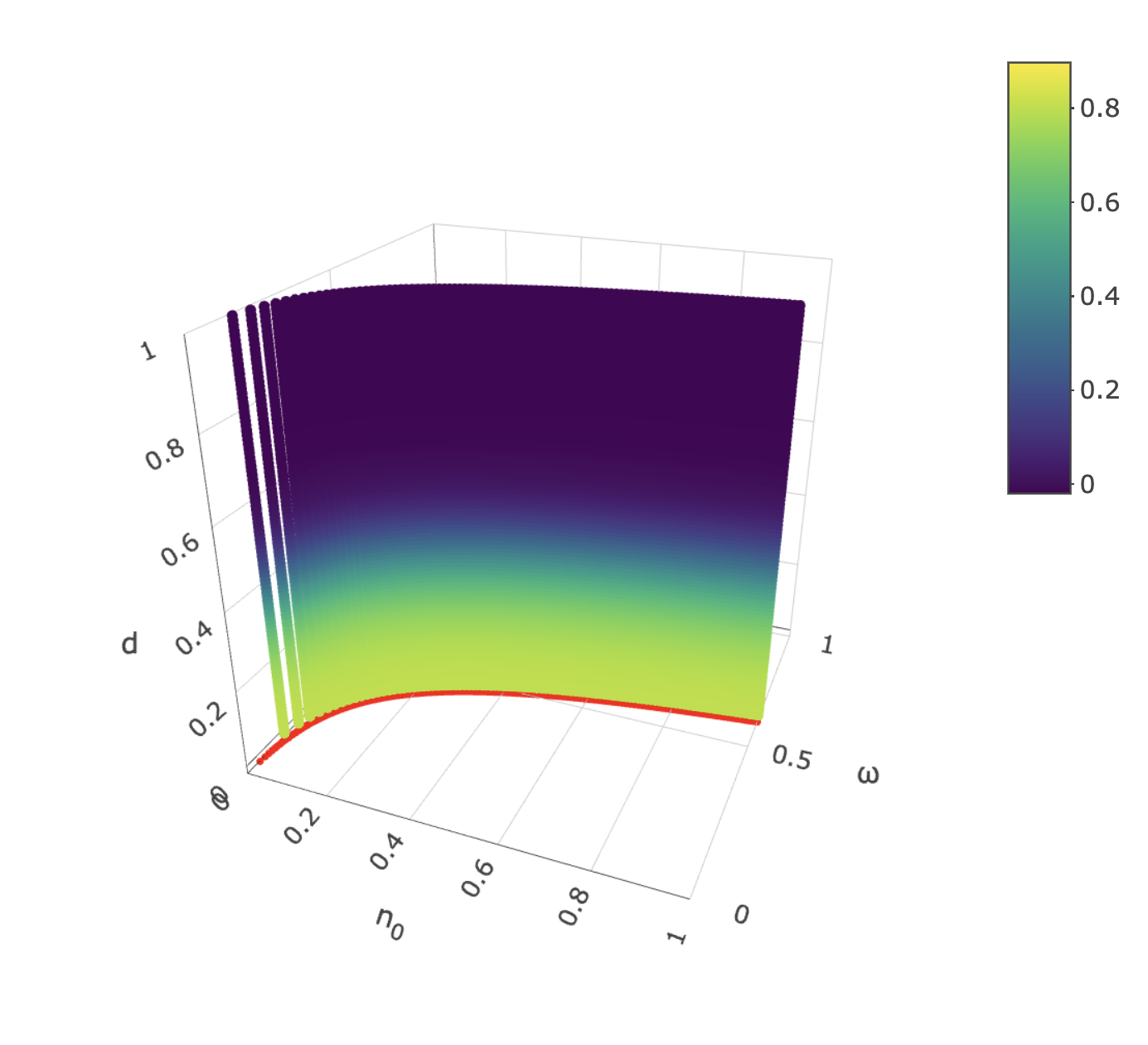}
        \label{p6}}
	\end{subfigure}\hfill
    \begin{subfigure}[$\beta^*=13.60$]{
		\centering
		\includegraphics[width=0.27\textwidth]{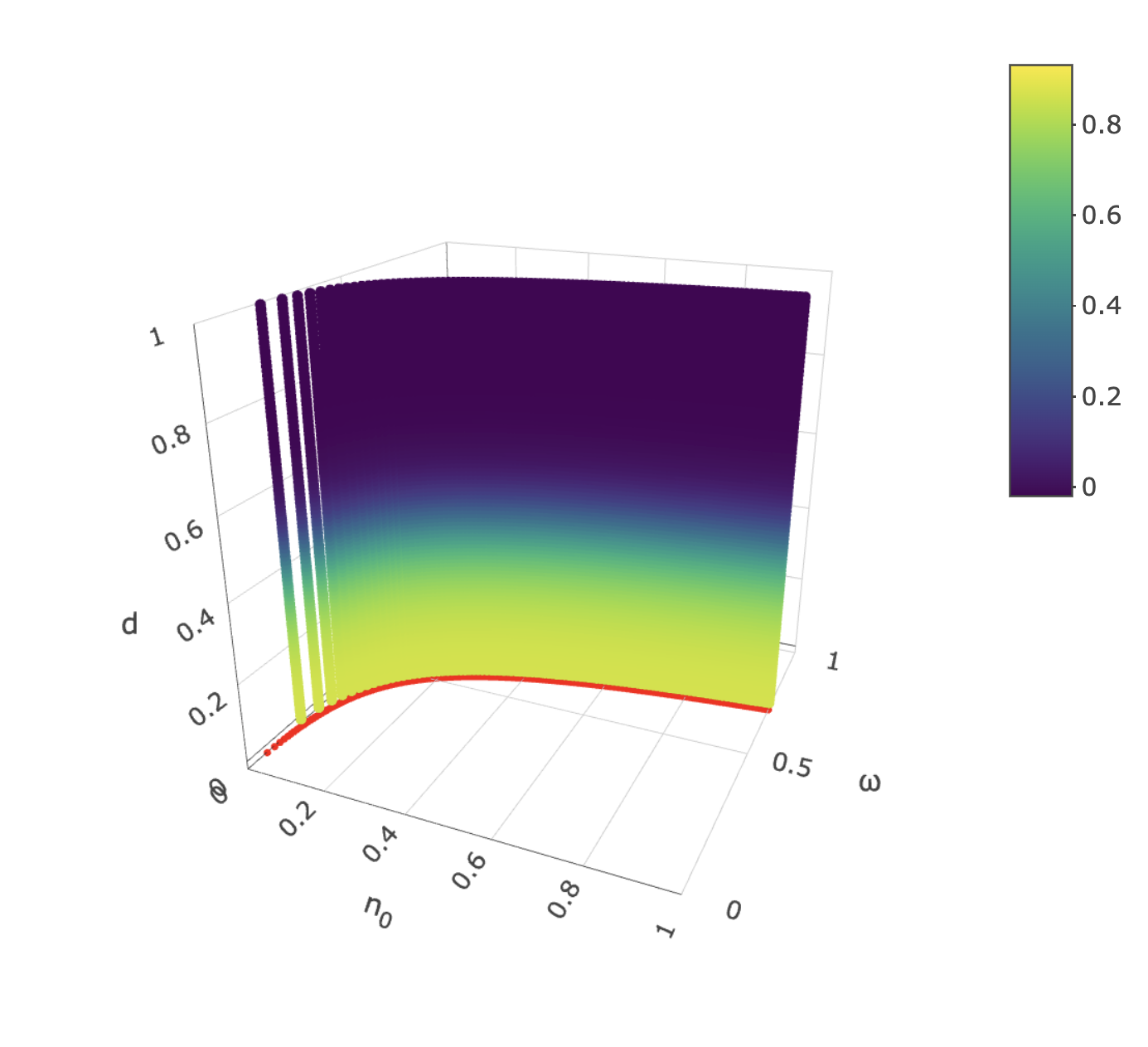}
        \label{p7}}
	\end{subfigure}
    \begin{subfigure}[$\beta^*=23.32$]{
		\centering
		\includegraphics[width=0.27\textwidth]{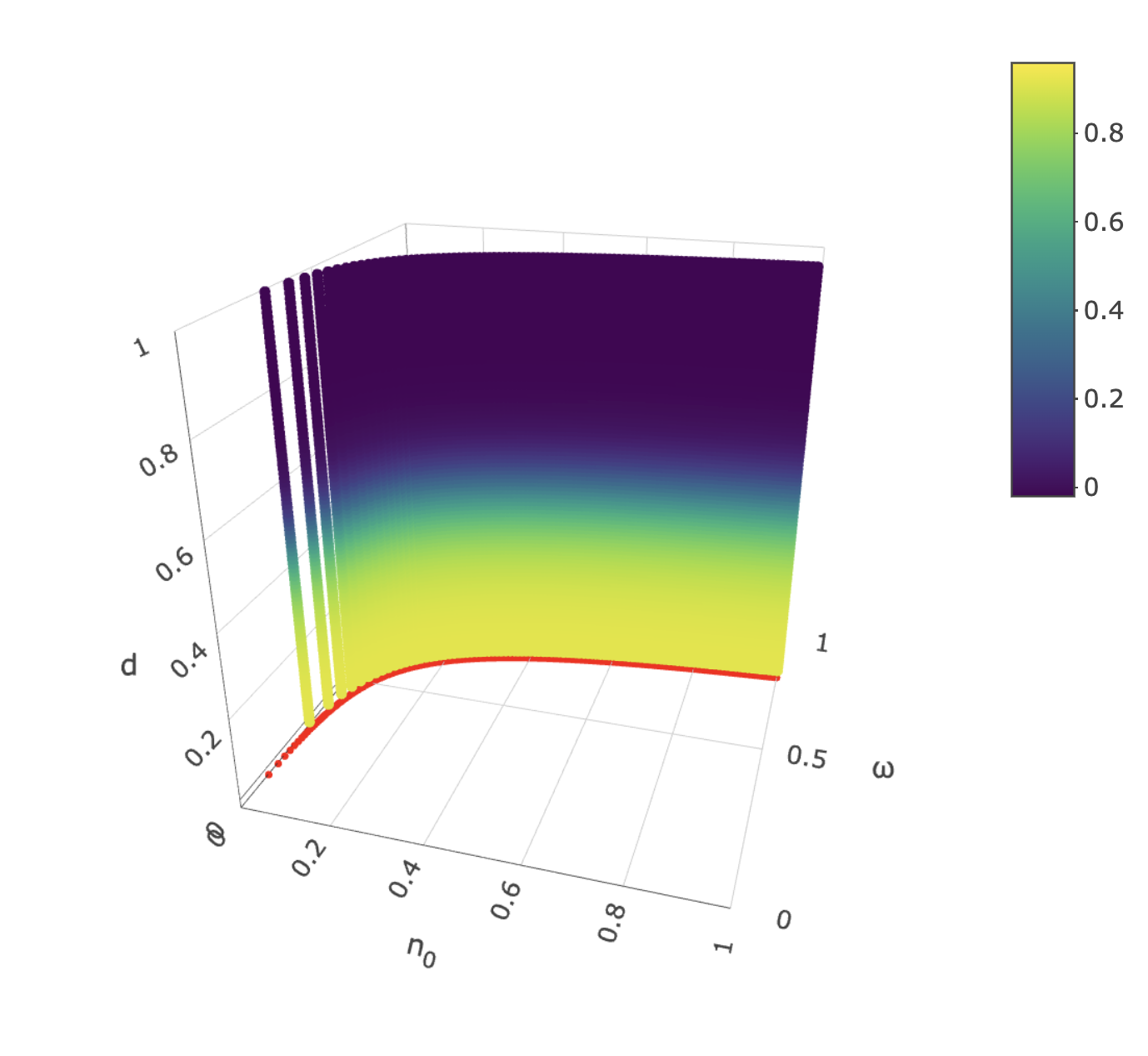}
        \label{p8}}
	\end{subfigure} \hfill
	\begin{subfigure}[$\beta^*=52.48$]{
		\centering
		\includegraphics[width=0.27\textwidth]{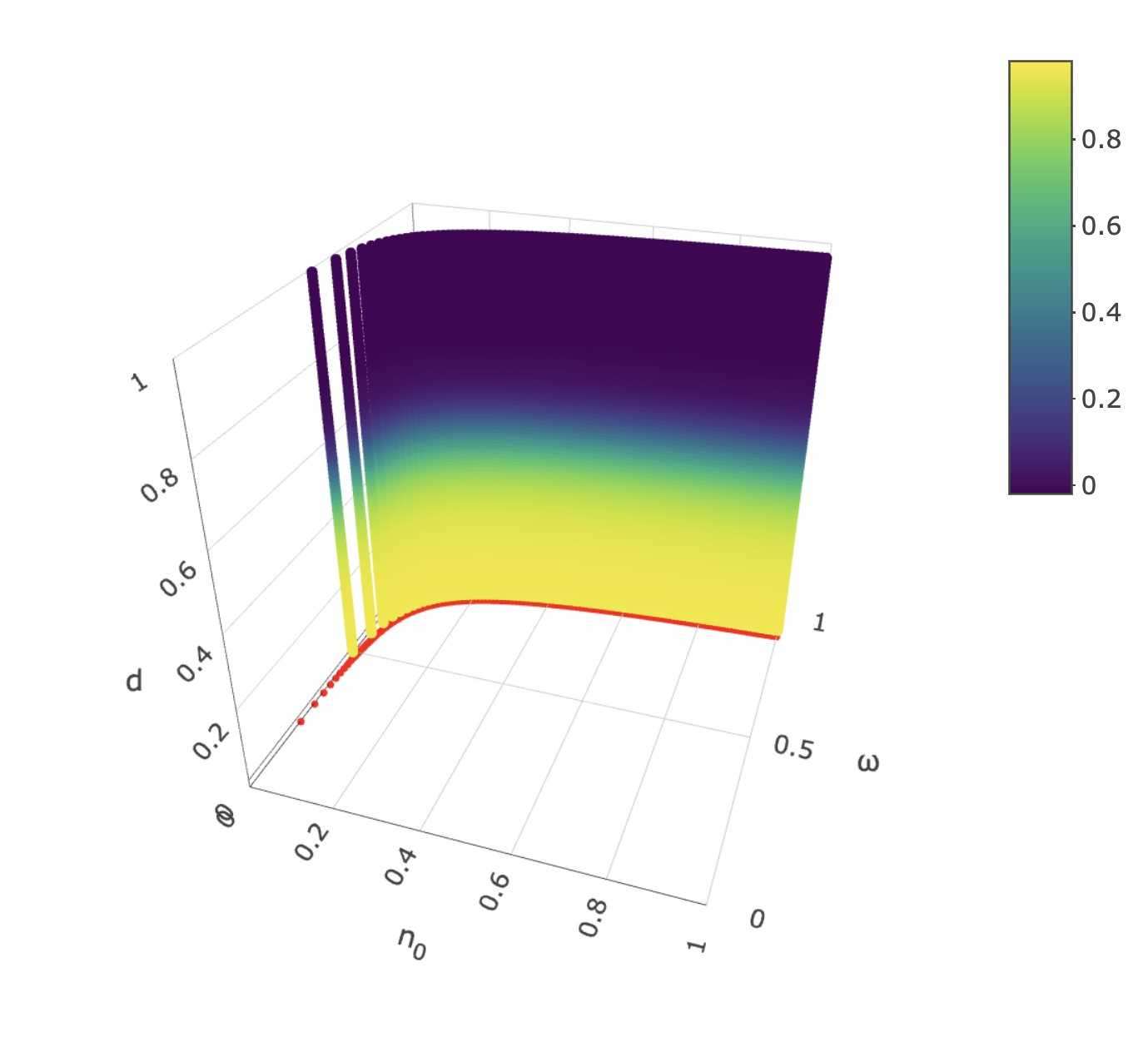}
        \label{p9}}
	\end{subfigure}\hfill
	\caption{Posterior weight $\Tilde{\omega}$ as a function of $n_0$, $\omega$ and $x_c$. Each panel represents all RMPs with a particular value of $\beta^{*}$}
    \label{equivalence.fig}
\end{figure}

\renewcommand{\thefigure}{S3}
\begin{figure}[h!]
   \centering
   \includegraphics[width=1\linewidth]{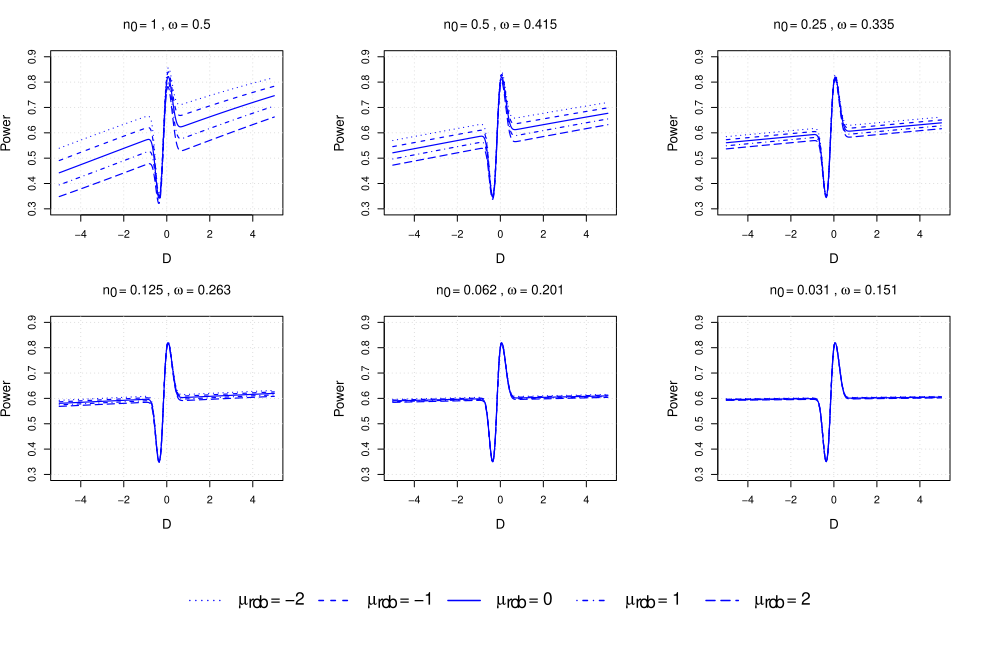}
   \caption{For each panel representing a different couples of $(\omega, n_0)$, power as a function of the prior-data conflict $D$ is displayed for five different values of the location of the robustification component of the RMP $\mu_{\text{rob}}$. Power is computed assuming a true mean treatment difference $\delta^{*}=0.31$.}
   \label{pow.loc}
\end{figure}

\clearpage

\renewcommand{\thefigure}{S4}
\thispagestyle{empty} % niente numero di pagina

\vspace*{\fill} % inizio centratura verticale
\begin{figure}[H]
   \centering
   \includegraphics[width=0.9\linewidth]{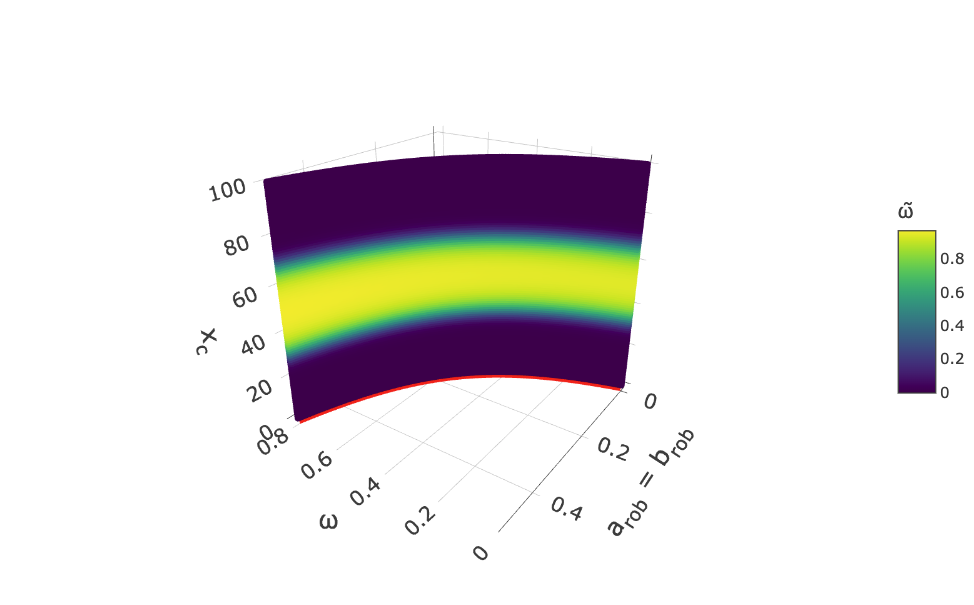}
   \caption{Posterior weight $\Tilde{\omega}$ as a function of $a_{\text{rob}}=a_{\text{rob}}$, $\omega$ and $x_c$. 
   The red curve in the horizontal plane represents all RMPs with $\beta^{*}=12.56$.}
   \label{3d_beta}
\end{figure}
\vspace*{\fill} % fine centratura verticale

\clearpage

\renewcommand{\thefigure}{S5}
\thispagestyle{empty}

\vspace*{\fill} % inizio centratura verticale
\begin{figure}[H]
    \centering

    % Immagine 1: Bias
    \includegraphics[width=0.3\textwidth]{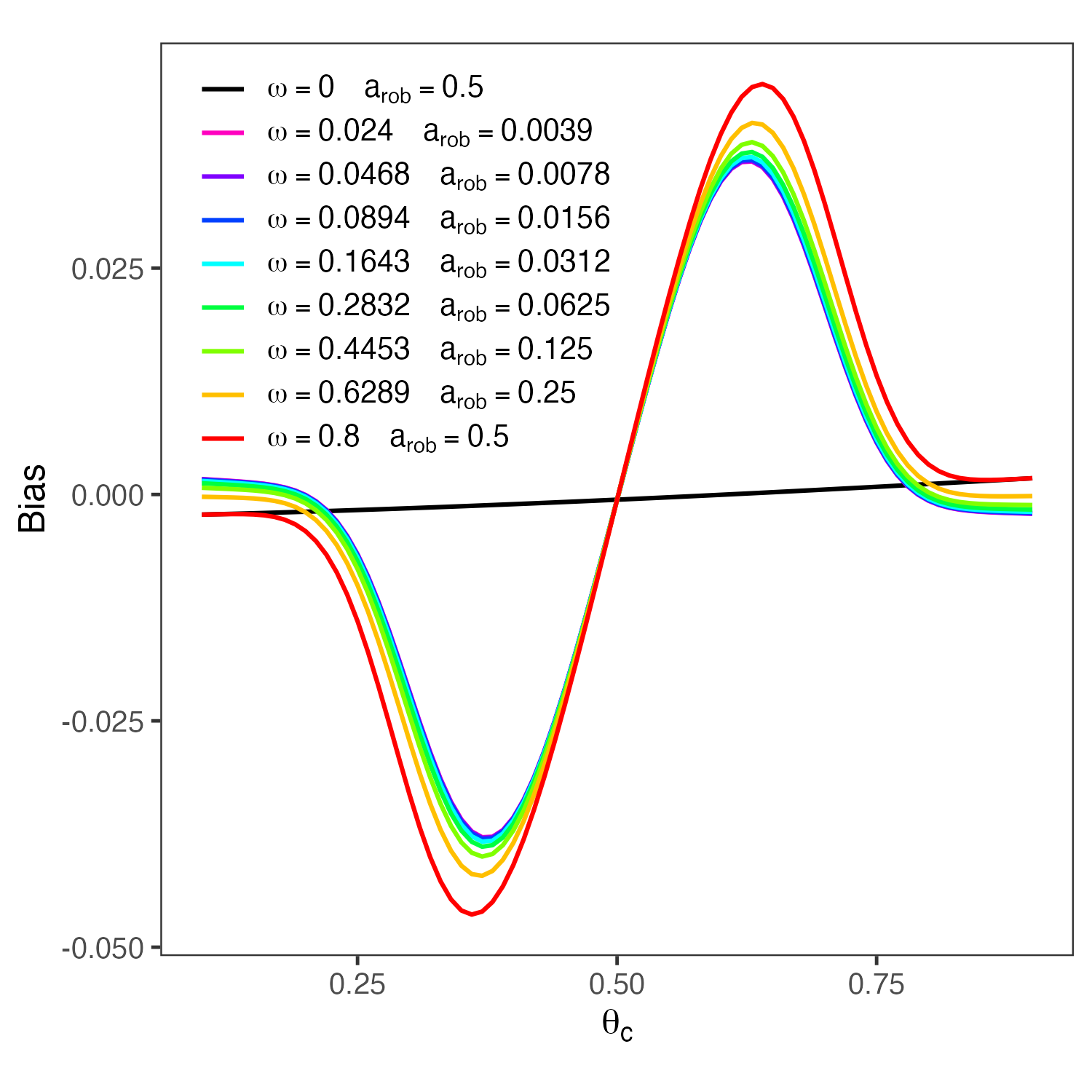}
    \hfill
    % Immagine 2: Variance
    \includegraphics[width=0.3\textwidth]{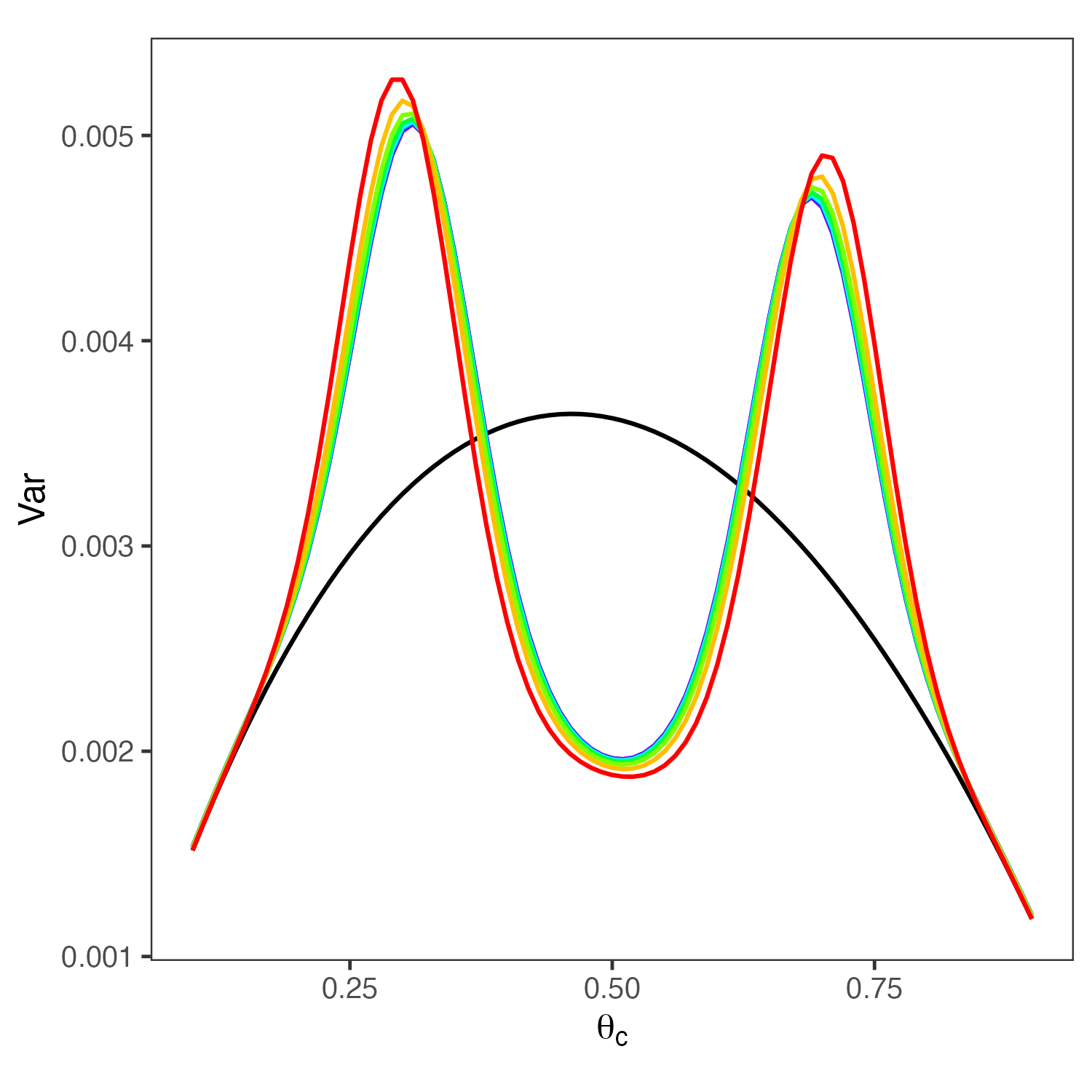}
    \hfill
    % Immagine 3: MSE
    \includegraphics[width=0.3\textwidth]{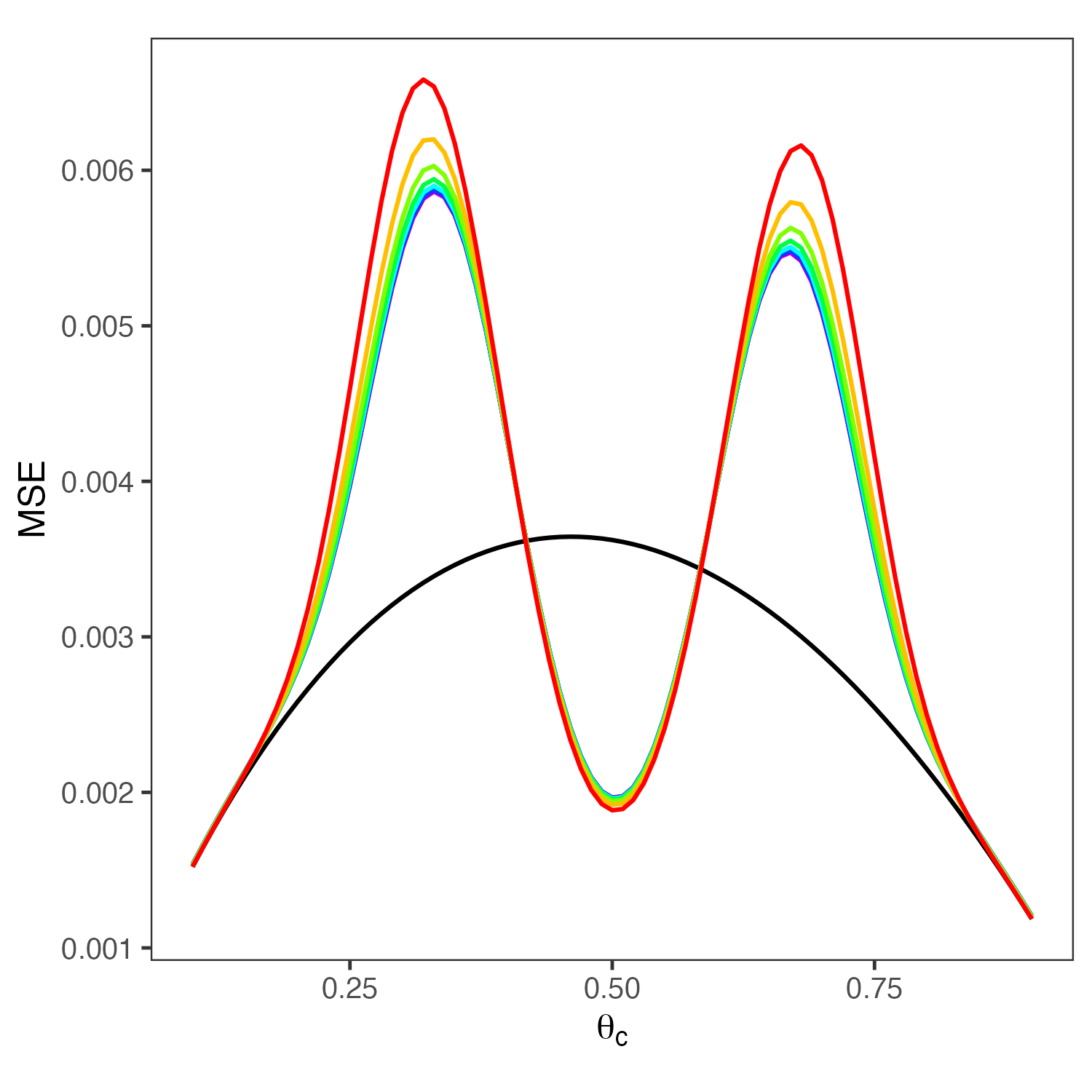}

    \caption{Panel~(a): bias; Panel~(b): variance; Panel~(c): mean squared error in the Beta--Binomial setting, all computed using the posterior mean of the treatment effect parameter~$\delta$ as the point estimate. Colors indicate different combinations of~$(\omega, a_{\text{rob}} = b_{\text{rob}})$, each corresponding to~$\beta^{*} = 12.56$.
}
    \label{S6.fig.beta}
\end{figure}
\vspace*{\fill} % fine centratura verticale

%\fi

\end{document}